\documentclass[12pt]{article}

\usepackage[labelsep=colon,labelfont={small,bf},singlelinecheck=on]{caption}
\usepackage[utf8]{inputenc}
\usepackage{amsmath,amssymb}
\usepackage{graphicx}
\usepackage{caption}
\usepackage{longtable}
\usepackage{multirow}
\usepackage{color}
\usepackage{dcolumn}
\usepackage{bm}
\usepackage[numbers,super,comma,sort&compress]{natbib}
\definecolor{background-color}{gray}{0.98}

\title{Electronic structure and non-linear optical properties of organic photovoltaic systems with potential applications on solar cell devices: A DFT approach}
\author{Alfredo Guillén-López\thanks{Instituto de Energías Renovables, Universidad Nacional Autónoma de México, Priv. Xochicalco s/n, Col.  Centro, Temixco, Morelos. CP 62580, Mexico}  \thanks{Cuerpo Acad\'emico de Energ\'ia y Sustentabilidad, Universidad Polit\'ecnica de Chiapas, Carretera Tuxtla-Villaflores km 1 + 500, Suchiapa, Chiapas, CP 29150, Mexico}, Miguel Robles$^{*}$, Jesús Muñiz$^{*}$\thanks{CONACYT-Universidad Nacional Aut\'onoma de M\'exico, Priv. Xochicalco s/n, Col.  Centro, Temixco, Morelos. CP 62580, Mexico}}

\begin{document}
\maketitle

\begin{abstract}

The use of eco-friendly materials for the environment has been addressed as a critical issue in the development of systems for renewable energy applications. In this regard, the investigation of organic photovoltaic (OPV) molecules for the implementation in solar cells, has become a subject of intense research in the last years. The present work is a systematic study at the B3LYP level of theory performed for a series of 50 OPV materials. Full geometry optimizations revealed that those systems with a twisted geometry are the most energetically stable. Nuclear independent Chemical shifts (NICS) values show a strong aromatic character along the series, indicating a possible polymerization in solid-state, via a $\pi - \pi$ stacking, which may be relevant in the design of a solar cell device. The absorption spectra in the series was also computed using Time Dependent DFT at the same level of theory, indicating that all spectra are red-shifted along the series. This is a promissory property that may be directly implemented in a photovoltaic material, since it is possible to absorb a larger range of visible light. The computed HOMO-LUMO gaps as a measurement of the band gap in semiconductors, show a reasonable agreement with those found in experiment, predicting candidate materials that may be directly used in photovoltaic applications. Non-linear optical (NLO) properties were also estimated with the aid of a PCBM molecule as a model of an acceptor, and a final set of optimal systems was identified as potential candidates to be implemented as photovoltaic materials. The methodological approach presented in this work may aid in the $in$ $silico$ assisted-design of OPV materials.

\end{abstract}

\clearpage


  \makeatletter
  \renewcommand\@biblabel[1]{#1.}
  \makeatother

\bibliographystyle{apsrev}

\renewcommand{\baselinestretch}{1.5}
\normalsize

\clearpage
\section*{\sffamily \Large Introduction}

The continuos global demand for energy consumption combined with the depletion of petroleum resourses and global warming effects, shows the urgent need of alternate sources for energy generation. The renewable energy is the best option, such as solar, wind, biomass and others. Sun energy reaching the Earth is many times larger than the present world energy consumption. In this respect, the solar photovoltaic cells are an excellent choice for the generation of electrical energy and they are also environmentally friendly\cite{Chaar11,Wild11,Solan11}.

Organic photovoltaic (OPV) materials and dye-sensitized solar cells (DSSCs) have been widely studied due to their potential as environment-friendly products, simple assembly technology, physical flexibility and low cost \cite{ORegan91,Sayama00}. The power conversion efficiency ( $\eta$ ) of OPV has achieved 10\% in 2012 \cite{Yang12,Li12,He11,He12}; while the DSSC has achieved the highest power conversion efficiencies, almost 12\% in 2013 \cite{Zhang13}. The OPV device typically consists of an organic bilayer or a bulk heterojunction (BHJ) composed of a donor (D) and an acceptor (A) material between two different electrodes. The operation of an OPV is somewhat different from that of an inorganic semiconductor based-PV cell. When a photon is absorbed in the donor material, an exciton (pair electron-hole) is created, this exciton can be separated at a D/A interface. When the electron is separated, it can be transfered to the acceptor material and transported to the cathode, the hole produced travels throughout the polymer and it is collected at the anode.

Development of an efficient BHJ polymer has become an active field of research, due to its low cost of fabrication and profitability for large-scale production. For many years, an electron-rich and electron-deficient moiety have been used to obtain low band gap polymers and with this methodology, new polymers have been developed. 

Further, Olivares-Amaya et al.\cite{Olivares11} found high-efficiency OPV materials within the Harvard Clean Energy Project initiative. The study was based upon 30 building blocks to obtain a set of 2.6 million molecules by using combinatorial molecule generation and strategies from drug discovery, pattern recognition and machine learning. On the other hand, a set of descriptors and 50 experimentally characterized molecules were used to parametrize a linear regresion that returns a final set of OPVs with desirable photovoltaic properties. The 50 systems experimentally characterized are polymeric structures with carbazole, thiophene and pyrrole derivates, and low band gap polymers such as benzothiadiazole, quinoxaline and thienopyrazine and others. 

In the literature, these materials were synthesized for organic thin film transistors and organic photovoltaic applications, but only a few have a theoretical study\cite{Blouin08,Jianhui:08,Mammo07}. Two of these systems were theoretically studied using Density Functional Theory (DFT) by Blouin et al. \cite{Blouin08}. In that work, it was obtained a correlation between DFT calculations and experimental HOMO, LUMO and band gap energies, revealing that the HOMO energy is fixed by the carbazole moiety, while the LUMO energy is related to the nature of the electron drawing comonomer. Furthermore, the band gap ($E_g$) in a set of the polymers is studied, ranging from 1.15 to 2.0 eV and absorb light from visible to near-infrared regions \cite{Jianhui:08,Mammo07}. The donor materials show three functional parts comprising an electron donor, electron  conductor and anchoring group (bridge) (D-B-A). In those studies, a DFT analysis states that to increase hole mobility in the polymer, to increase molecular weight and to decrease polydispersity, the enlargement of the $\pi -\pi$ bridge in the polymer or the introduction of planar monomers, should be applied.

Theoretical studies in this area have increased in recent years due to high performance computing and optimization of computational chemistry codes. The theoretical approach is the best tool to overcome the challenges in the experimental synthesis and to explore alternatives that reduce cost of materials production and processing. The use of DFT represents a reliable alternative to tackle these tasks, since the methodology explores electronic structure and spectroscopic properties on this type of materials. In this work we calculated the electronic structure, nonlinear optical properties and optical properties of the 50 experimentally characterized molecules, as described above by using DFT and Time-Dependent DFT.

The quest of novel OPV materials via $in$ $silico$ techniques has been performed with certain dependability. Nevertheless, no comprehensive study on the excited state and nonlinear optical properties is available in open literature. The aim of this work is to give insights into the design of OPV materials for solar cell devices through the prediction of nonlinear optical parameters and the physical behavior in the excited state, on the series of 50 OPV systems experimentally characterized. The anticipated evaluation of these parameters represents a new methodology to accelerate the synthesis and development of renewable energy materials.

\section*{\sffamily \Large Computational Details}


In the present work, the ground state geometries and excited states of fifty organic photovoltaic materials were performed with TURBOMOLE computational code \cite{program}. For the ground state properties, we used the three-parameter functional of Becke, and the correlation of Lee, Yang and Parr (B3LYP) \cite{Lee88,Becke93}. The def2-TZVP basis set on all atoms \cite{Weigend98} was used due to a better quality than other standard basis sets. For the excited states properties such as UV-vis spectra, oscillator strengths and Light Harvesting Efficiencies, we used the TD-DFT methodology\cite{Casida:95} on the optimized geometries with the same functional and basis set than those used at the ground state.

According to the results, the UV-vis absortion spectra were simulated by Gaussian functions \cite{Stephens10}, using the TmoleX display program \cite{Steffen10}, calculated from the first ten excited states at the ground state geometries.

The power conversion efficiency ($\eta$) for the overall solar cell is determined from the following equation:
\begin{equation}\label{eficiencia}
\eta = \dfrac{J_{SC} V_{OC} FF}{P_{in}},
\end{equation}
where $J_{SC}$ is the photocurrent density, $V_{OC}$ is the open circuit voltage, $FF$ is the fill factor and $P_{in}$ is the power density of the incident light. $P_{in}$ was fixed to conditions of laboratory, that is, a standard value of incident spectrum at AM1.5G with an intensity of 1000 W/m$^2$ (100 mW/cm$^2$) at room temperature.

The $V_{OC}$ values can be aproximated  from the following equation:
\begin{equation}\label{Voc_aprox}
V_{OC} \approx \dfrac{n \ k_{B} \ T}{e} \ln \left ( \dfrac{J_{SC}}{J_0}  \right ) ,
\end{equation}
where $n$ is the electron density, $k_B$ is the Boltzmann constant, $T$ is the temperature, $J_0$ is the inverse saturation current and $e$ is the elemental charge of electron \cite{Kippelen09, Qi13}. For this work, we approached the open circuit voltage with the model of Scharber \cite{Scharber06}, where the energy of the frontier molecular orbitals (HOMO and LUMO) are considered for the donor and the acceptor, respectively. This is expressed in the following equation:

\begin{equation}\label{ecu-Voc}
V_{OC}=\left ( \dfrac{1}{e} \right) \left | E_{H}^{D} - E_{L}^{A} \right | - 0.3 V,
\end{equation}
where $E_H^D$ is the HOMO energy of the donor, $E_L^A$ is the LUMO energy of the acceptor, $e$ is the elemental charge of electron and the 0.3V value is the loss factor related to the heterojunction design. The equation \ref{ecu-Voc} is used more frequently used in these calculations\cite{Muhlbacher06, Dongmei14,Bourass13}.

Short circuit current $J_{SC}$ values were estimated with the following equation:
\begin{equation}\label{circuit_integral}
J_{SC} = \int\limits_{\lambda} LHE(\lambda)\phi_{in} \eta_{collec} d \lambda ,
\end{equation}
where $LHE(\lambda)$ is the light-harvesting efficiency at $\lambda$, $\phi_{in}$ is the electron injection efficiency and $\eta_{collec}$ is the electron collection efficiency \cite{Ding13,Zhang12}.

We assume that the $\eta_{collec}$ value is constant. To obtain large values of $J_{SC}$ we need large values of LHE and $\phi_{in}$. LHE was obtained from:
\begin{equation}\label{LHE}
LHE(\lambda) = 1 - 10^{-f},
\end{equation}
where $f$ is the oscillator strength of the donor material associated with the wavelength of maximum absortion $\lambda_{max}$ \cite{Fan12,Zhang12-1}. $\phi_{in}$ is related to the force of electron injection ($\Delta G_{inj}$) from the excited state of the donor to the LUMO of the acceptor. We approached $\Delta G_{inj}$ as follows \cite{Wang11}:
\begin{equation}\label{delta_inye}
\Delta G_{inj} = E_{ox}^D - E_{0-0}^D - E_L^A ,
\end{equation}
where $E_{ox}^D$ is the ionization potential of the ground state of the donor, $E_{0-0}^D$ is the first excited state energy of the donor  \cite{Wichien12,Ryuzi04}. As previously suggested, $E_{L}^{A}$ is the LUMO energy of an acceptor. According to Scharber’s model \cite{Scharber06}, we have considered the PCBM system as the acceptor. Furthermore, we computed its LUMO energy from the experimental geometry at the PBE/QZVP \cite{Perdew:96} level of theory in order to finely reproduce the experimental value of -4.3 eV. We found that  $E_{L}^{A} = -4.026 $ eV, in agreement with the standard value found in laboratory. We used this theoretical result in the computations of $V_{OC}$ and $\Delta G_{inj}$, as given by Equations \ref{ecu-Voc} and \ref{delta_inye}, respectively. On the other hand, the ionization potential was calculated from 
\begin{equation}
E_{ox}^D = E_0 - E_0^+,
\end{equation}
here $E_0$ is the ground state energy and $E_0^+$ is the energy of the cation\cite{Lewars11}.

The nonlinear optical properties, such as isotropic polarizability $\alpha$, anisotropy of the polarizability  $\Delta \alpha$ and first order  hyperpolarizability $\beta_{tot}$, were calculated with the following equations \cite{Zhang10,Kosar11,Mohr15}:

\begin{equation}\label{alfa}
\alpha = \dfrac{\alpha_{xx} + \alpha_{yy} + \alpha_{zz} }{3},
\end{equation}

\begin{equation}\label{delta_alfa}
\Delta\alpha =  \sqrt{ \dfrac{ (\alpha_{xx} - \alpha_{yy} )^2 + (\alpha_{yy} - \alpha_{zz} )^2 + (\alpha_{zz} - \alpha_{xx} )^2   }{2} },
\end{equation}

\begin{equation}\label{beta}
\beta_{tot} = \sqrt{( \beta_{xxx} + \beta_{xyy} + \beta_{xzz} )^2 + (\beta_{yyy} + \beta_{xxy} + \beta_{yzz} )^2 + ( \beta_{zzz} + \beta_{xxz} + \beta_{zyy} )^2},
\end{equation}

where $\alpha_{ii},\  \beta_{xxi}, \ \beta_{iyy}$ and $\beta_{izz}$ are tensor components of polarizability and hyperpolarizability, respectively with $i=x, y, z$.  $E=0.01 $ a.u was used\cite{Xie15}.



\section*{\sffamily \Large Results and discussion}


\section*{\sffamily \Large Structural description of OPV systems}
 
\begin{figure}
\begin{center}
\includegraphics[width=0.8\columnwidth,keepaspectratio=true]{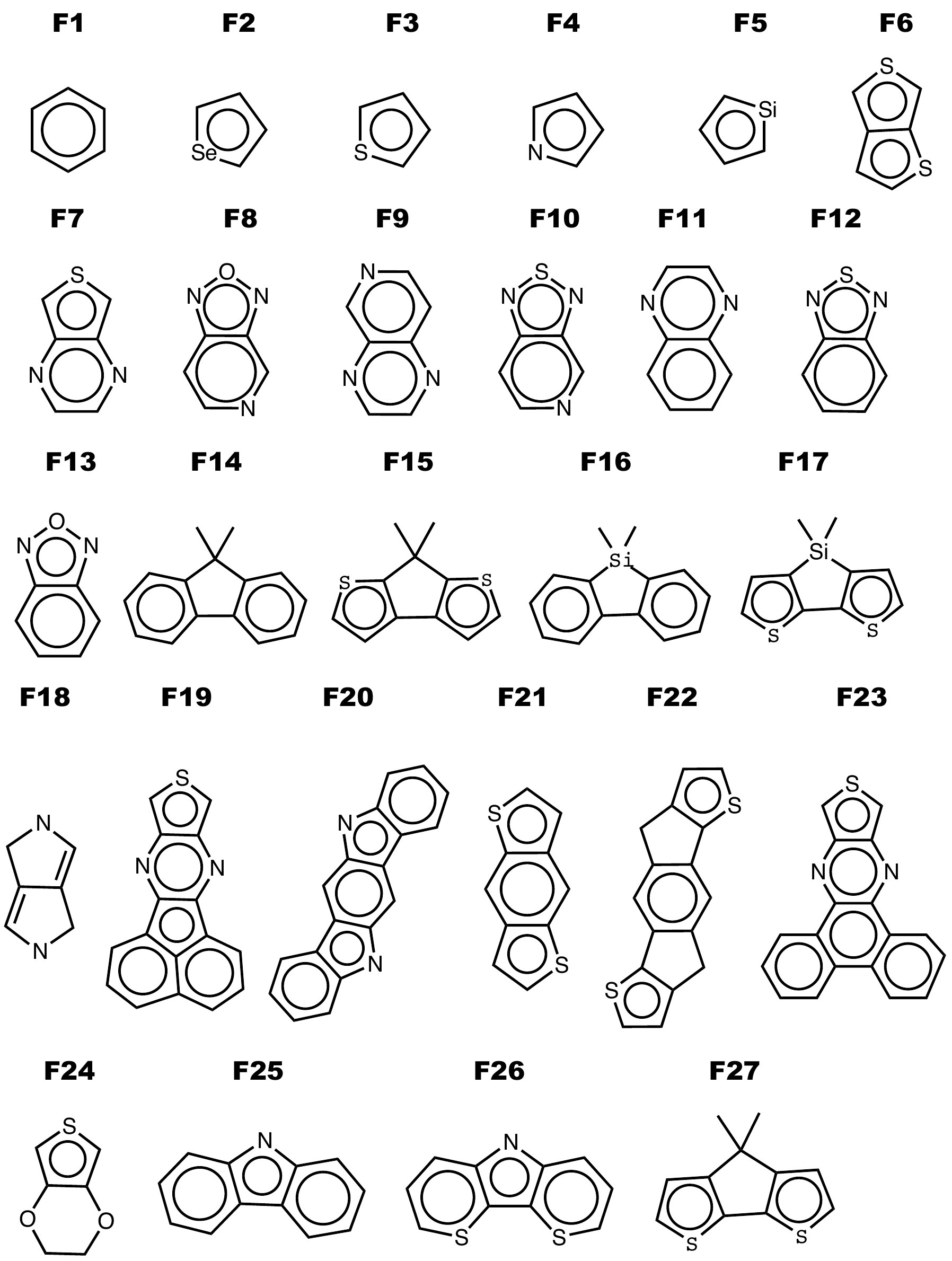}
\end{center}
\caption{\label{moietF1}Schematic representation of building blocks inside the 50 molecular systems.}
\end{figure}

The present study is based upon the calculation of electronic structure properties of a series of 50 organic materials, which were experimentally characterized \cite{Olivares11,Blouin08,Chen09,Mondal09,Muhlbacher06,Wang:08,Yuen10}. This set of molecules is presented in Fig. {\bf S.1-S.3 } of SI and it is used in our work to give insights into relevant electronic structure properties that may be of special interest in the development of novel photovoltaic materials. Such systems will be identified from S1 to S50 throughout this work. The set of systems under study is basically formed by a subset of 27 fragments, as shown in Fig. \ref{moietF1}. This group of molecules will be referred to as a fragment F1 to fragment F27 along the present study.

It is important to highlight that inside the main group of molecules, there are systems with exactly the same stoichiometry. That is, S6=S8, S10=S11, S3=S16=S18, S46=S48. This is the case, since the aliphatic chains were suppressed in the systems to speed-up the computational time\cite{Olivares11}. We also considered the notation ($\sim$) to denote those systems that share some structural coincidences, which are detailed in Table {\bf S.1-S.2} of SI.

\begin{figure}
\begin{center}
\includegraphics[trim= 0.8in 1in 0.5in 1in,width=0.5\columnwidth,keepaspectratio=true,angle=-90]{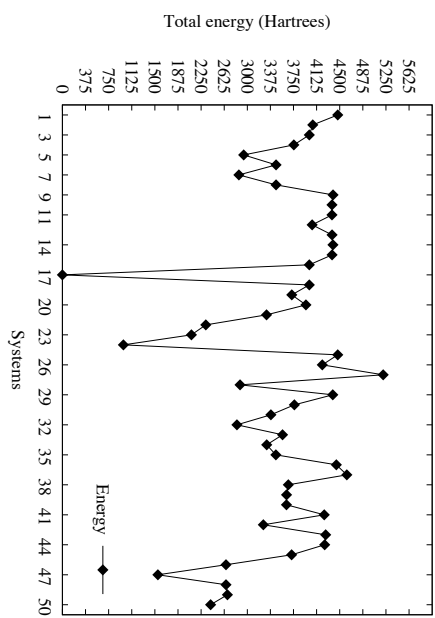}
\end{center}
\caption{\label{enerbase}Total energy of the 50 molecular systems normalized with respect to S17}
\end{figure}

The 50 systems inside the training set were fully optimized according to the methodology presented in the Computational Details. The total energies of the systems under study are presented in Fig. \ref{enerbase} with respect to the lowest energy systems in the series, which corresponds to the S17 molecular system. The systems S17, S23, S24 and S47 represent the most stable configurations in the group. System S17 contains fragment F2, while the most unstable, S27 system, presents two F3 fragment units. It is important to highlight that the ground state geometries are not necessarily planar and in some cases, different angles of torsion located at the fragments are observed (see Fig. {\bf S.4-S.10} of SI).

\begin{figure}
\begin{center}
\quad {\bf a) } S1 \qquad\qquad \qquad\qquad \qquad\qquad\quad  {\bf b) } S13
\vspace{0.3cm}

\includegraphics[width=2.3in]{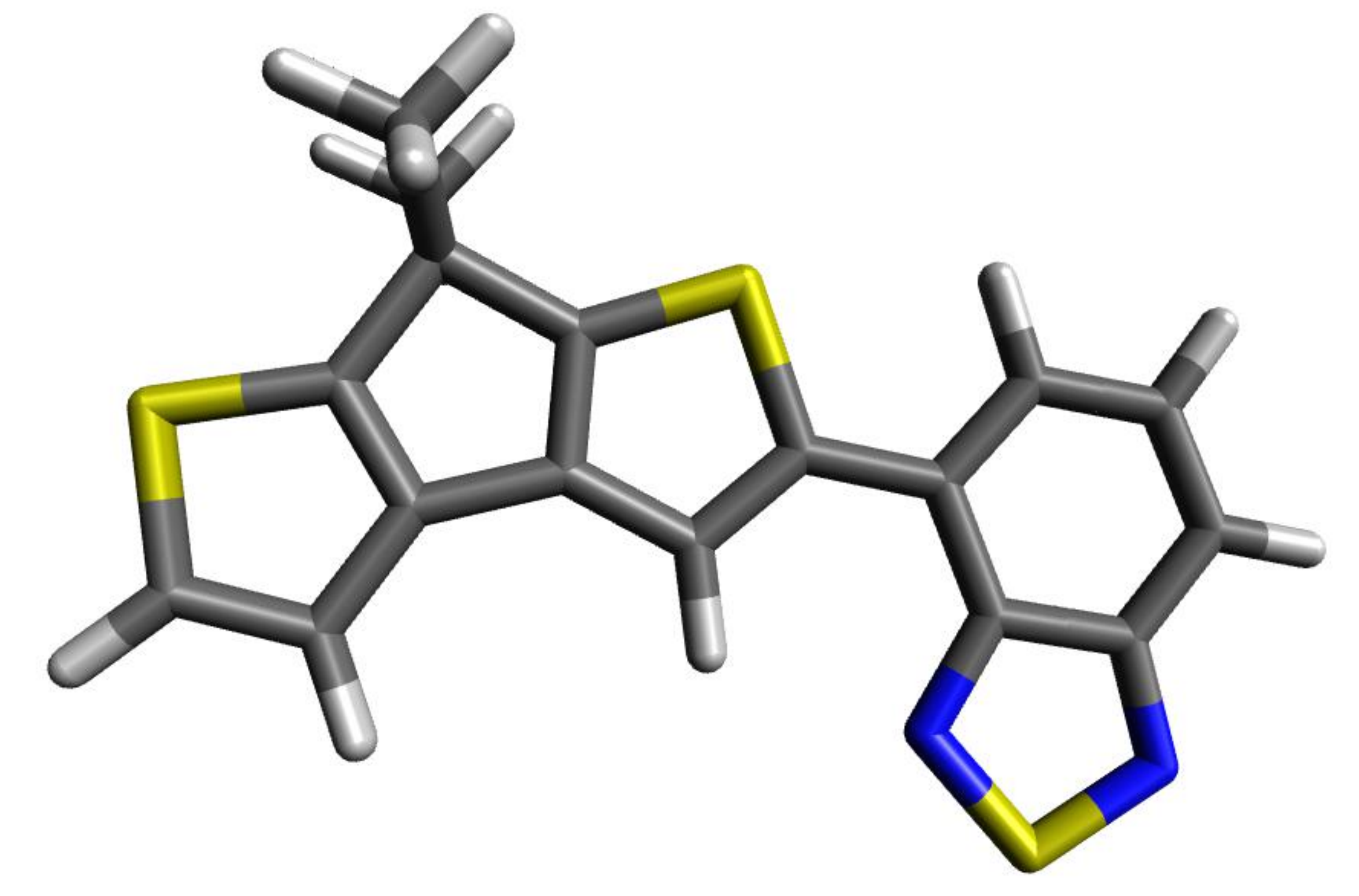} \ \ \includegraphics[width=3.2in]{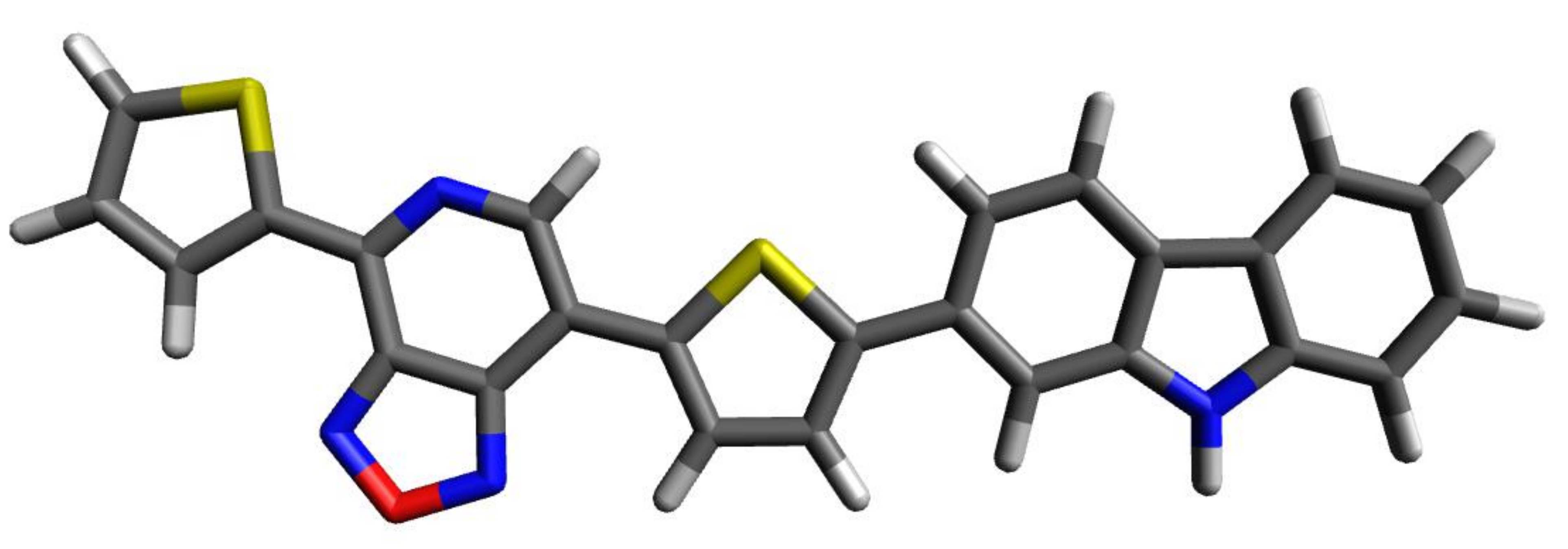} 
\vspace{1cm}

\quad  {\bf c) } S21 \qquad\qquad \qquad\qquad \qquad\qquad\quad  {\bf d) }S25
\vspace{0.3cm}

\includegraphics[width=3.1in]{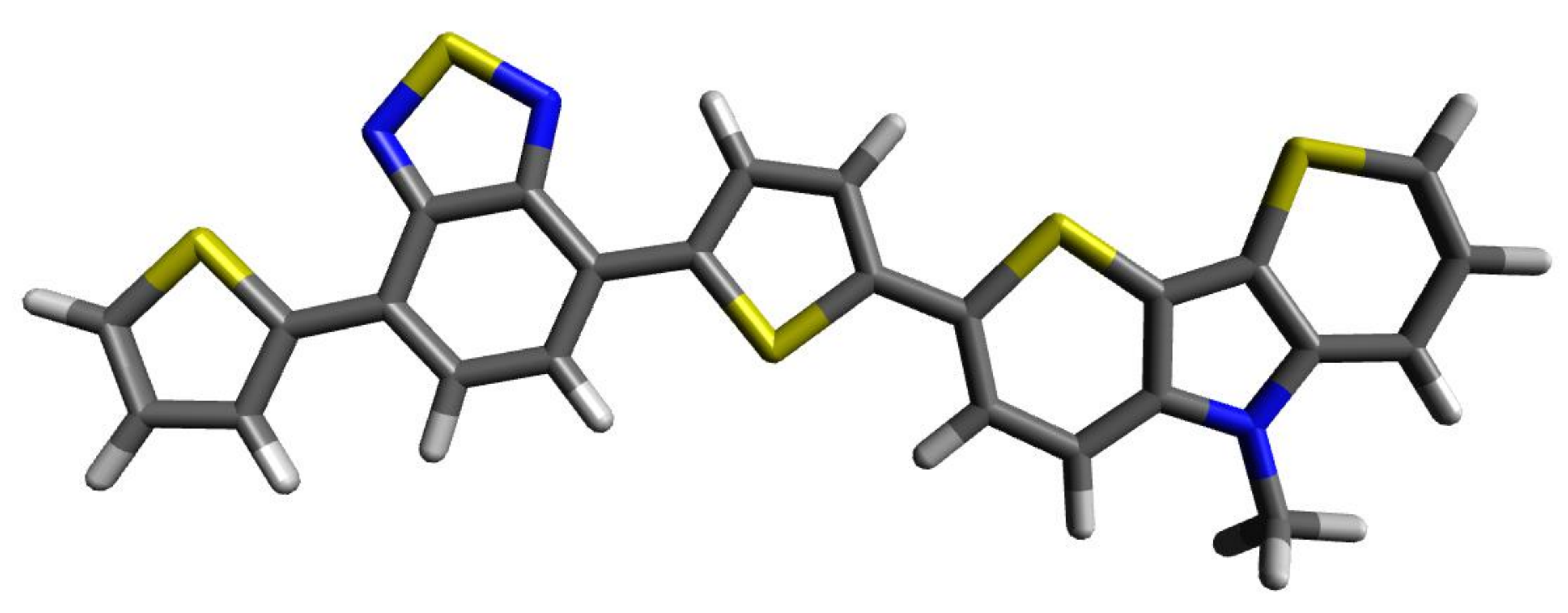} \ \includegraphics[width=2.3in]{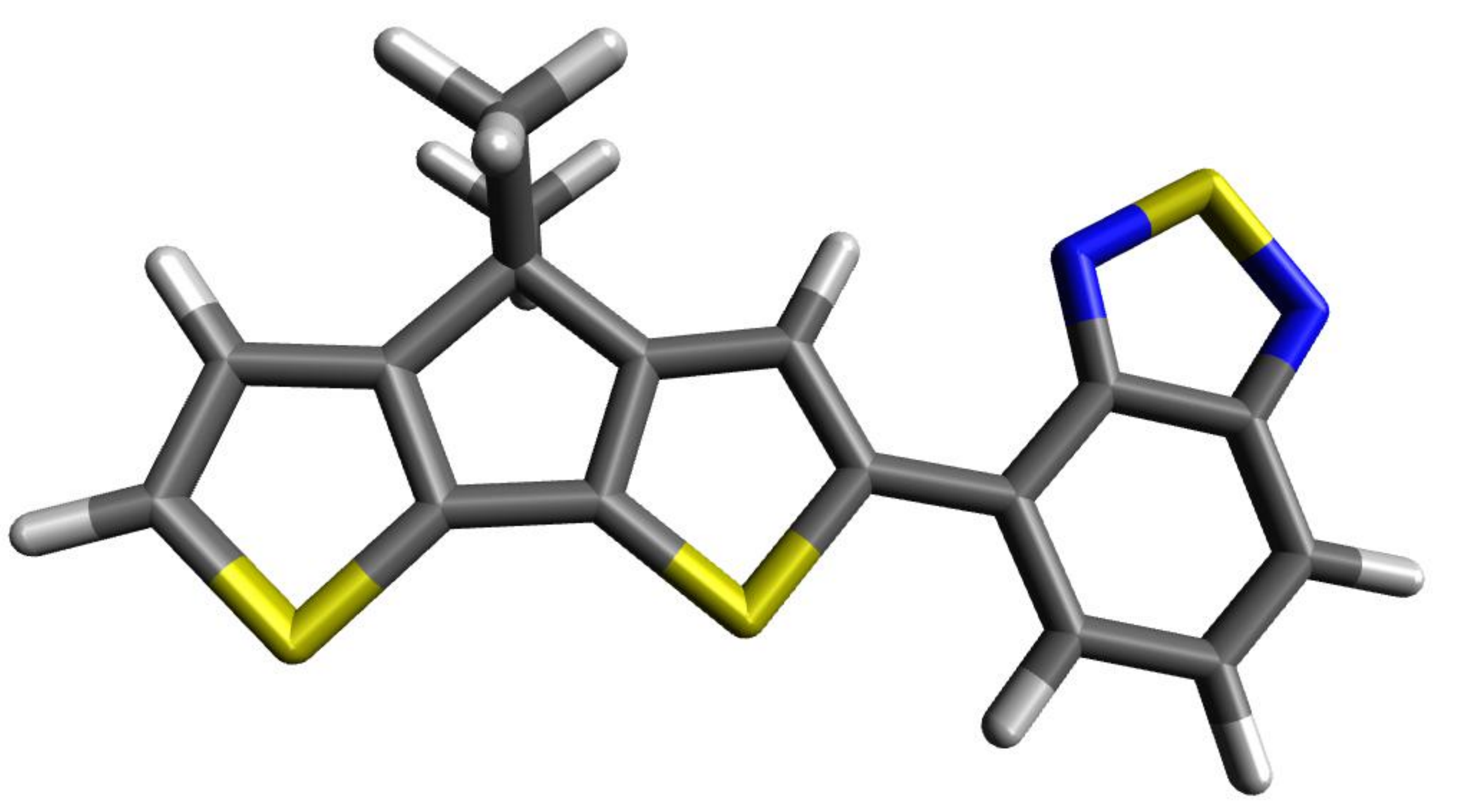}
\vspace{1cm}

\quad  {\bf e) } S26 \qquad\qquad \qquad\qquad \qquad\qquad\quad  {\bf f) } S33
\vspace{0.3cm}

 \includegraphics[width=2.3in]{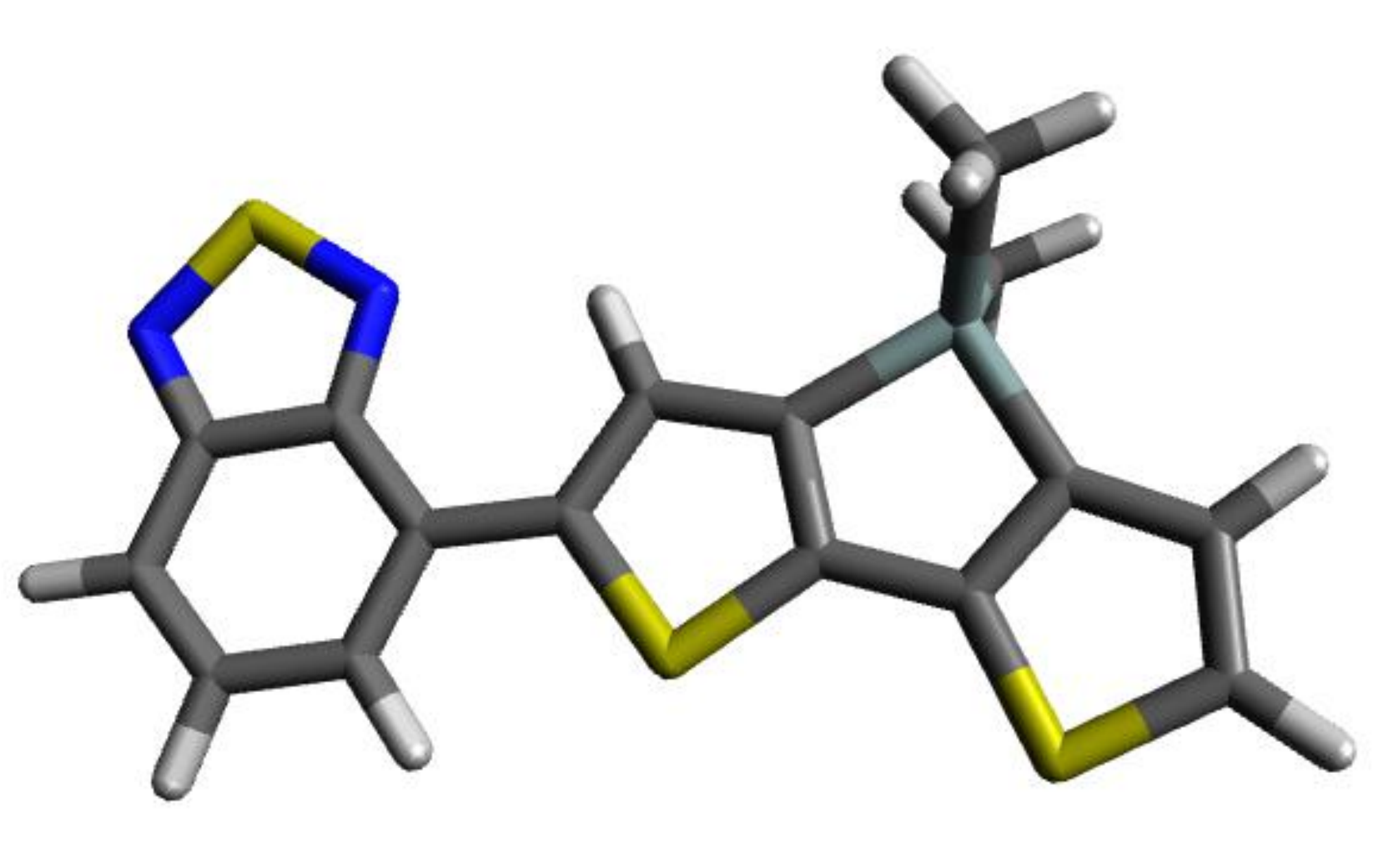} \ \includegraphics[width=3.2in]{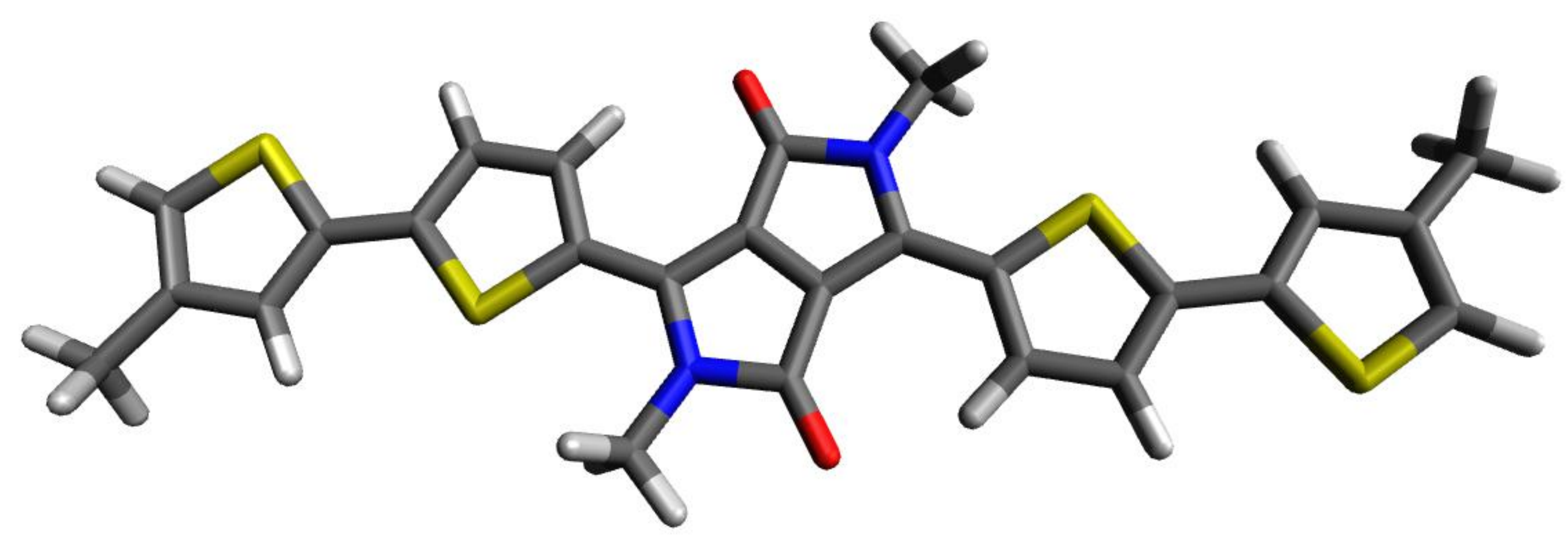}
\end{center}
\caption{\label{planarsys}Molecular systems with planar geometry: a) S1, b) S13, c) S21, d) S25, e) S26, f) S33.}
\end{figure}

\begin{figure}
\begin{center}
\quad {\bf a) } S8   \qquad  \qquad  \qquad \qquad\qquad \qquad\qquad {\bf b) } S31
\vspace{0.35cm}

\includegraphics[width=3.2in]{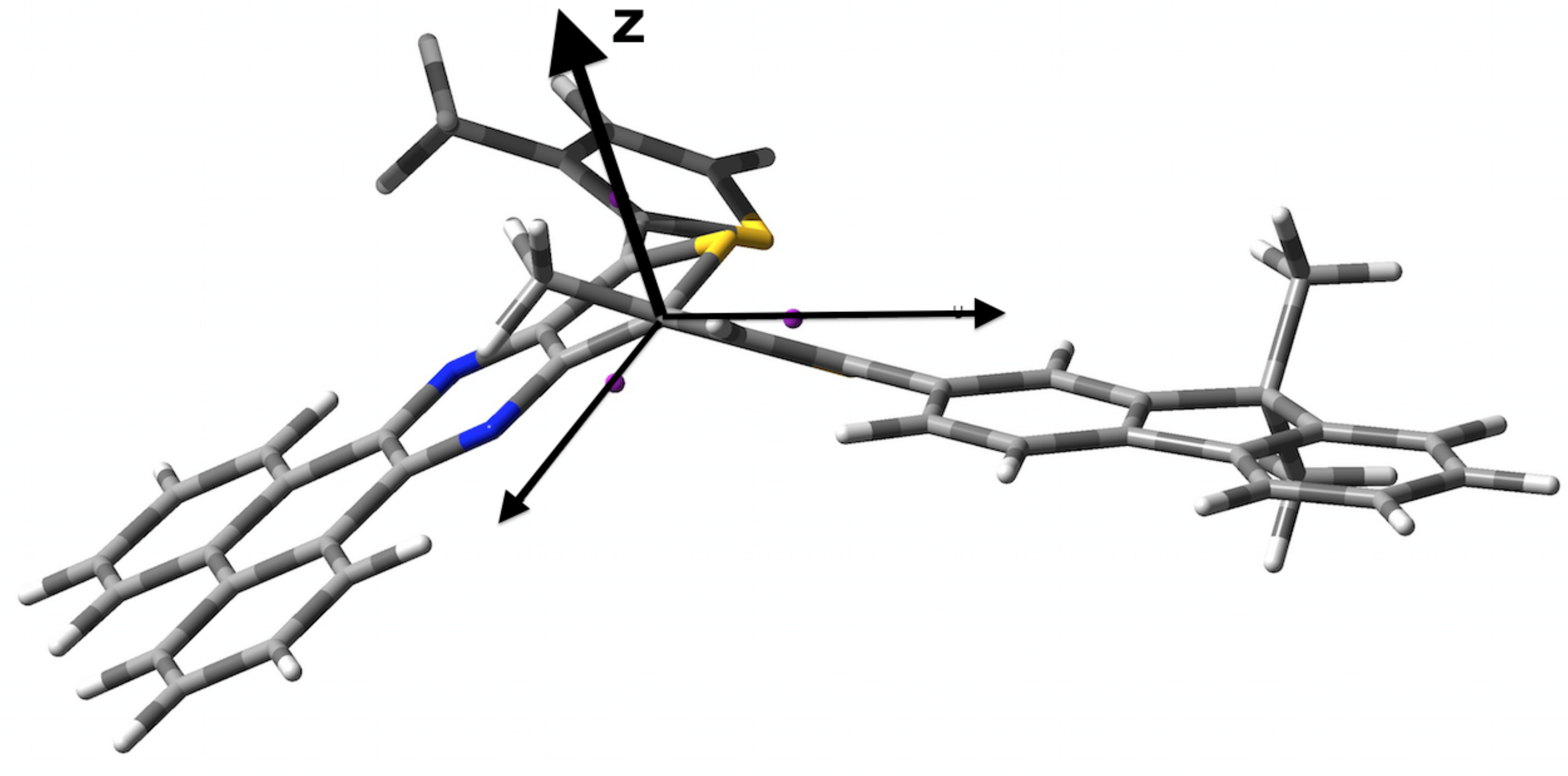} \  \ \includegraphics[width=2.2in]{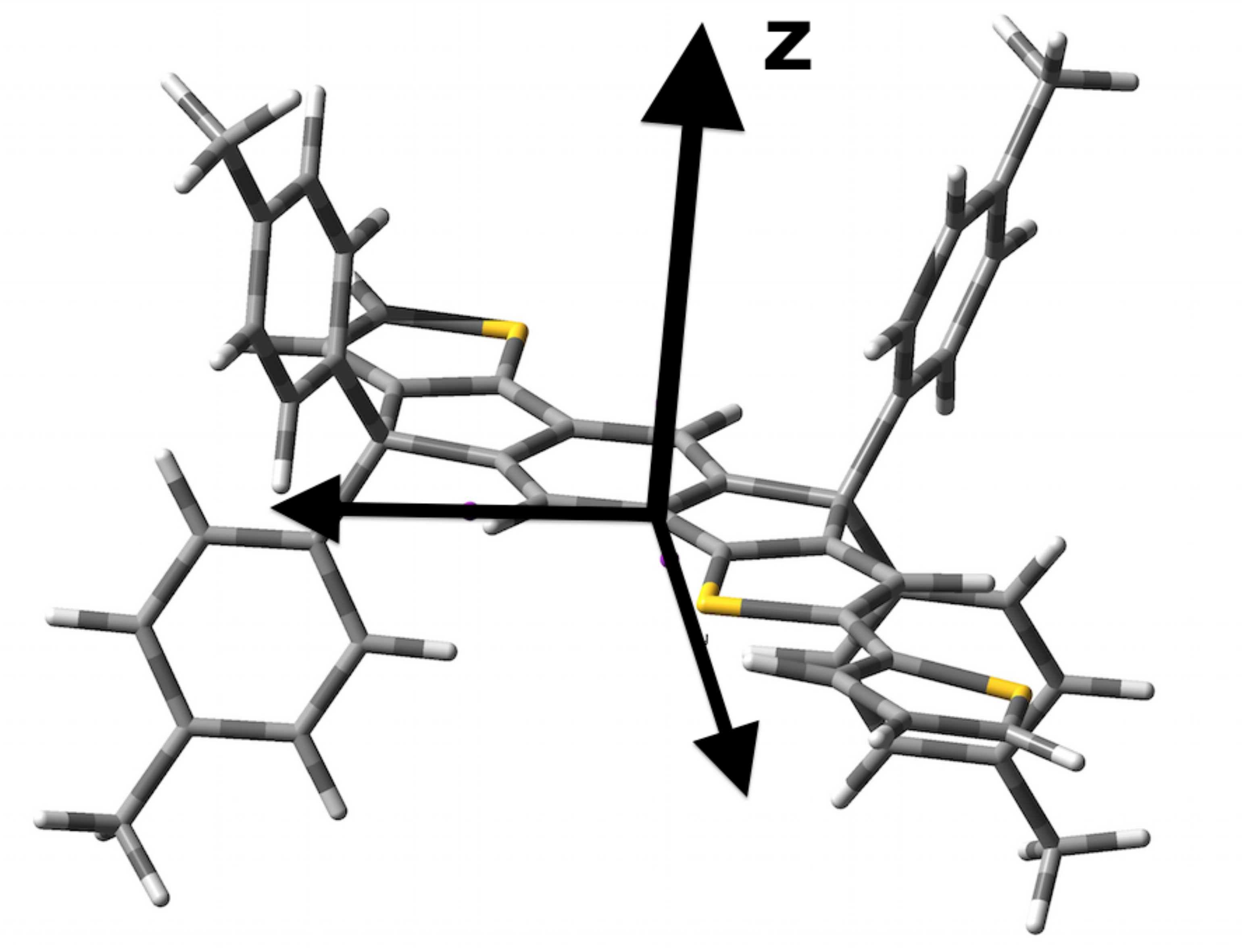}
\vspace{0.35cm}

\quad  {\bf c) } S3   \qquad  \qquad  \qquad \qquad\qquad \qquad\qquad {\bf d) } S27
\vspace{0.35cm}

\includegraphics[width=3.4in]{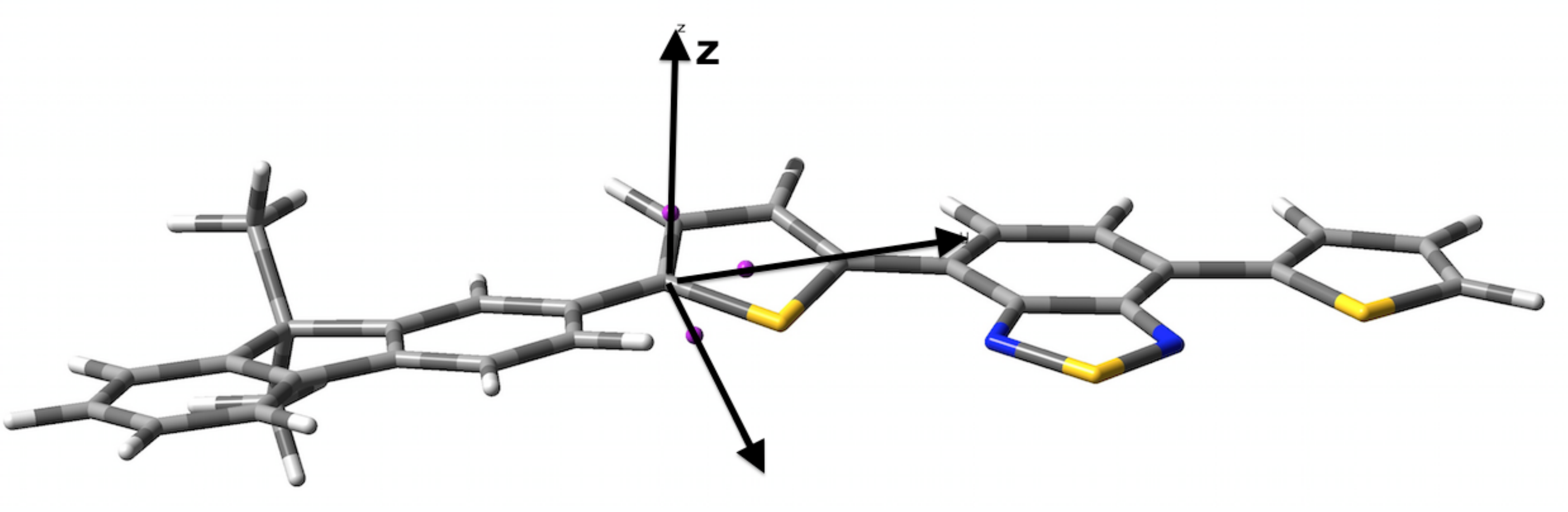} \  \includegraphics[width=2in]{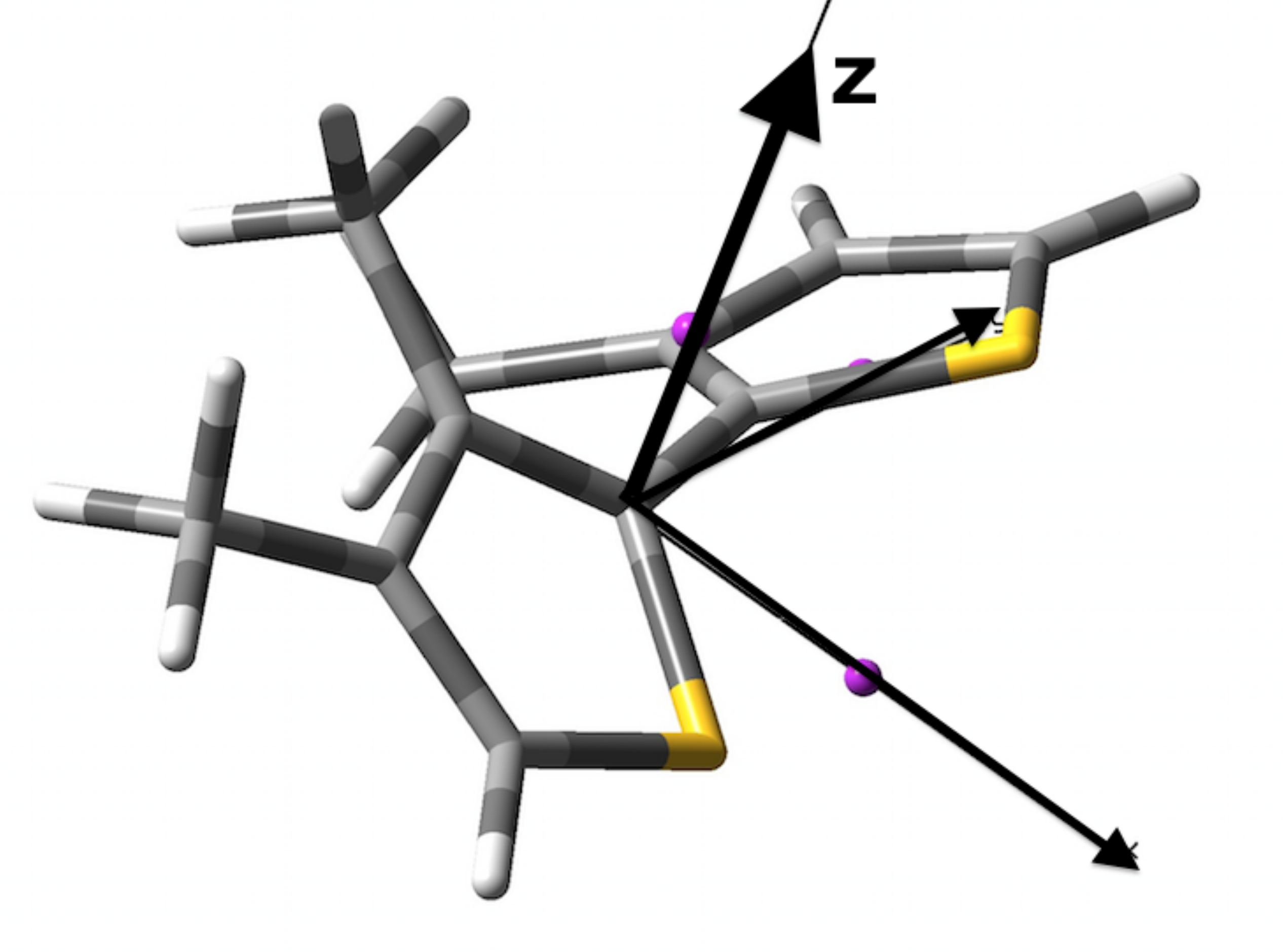}
\vspace{0.35cm}

\quad  {\bf e) } S47
\vspace{0.15cm}

 \includegraphics[width=5.1in]{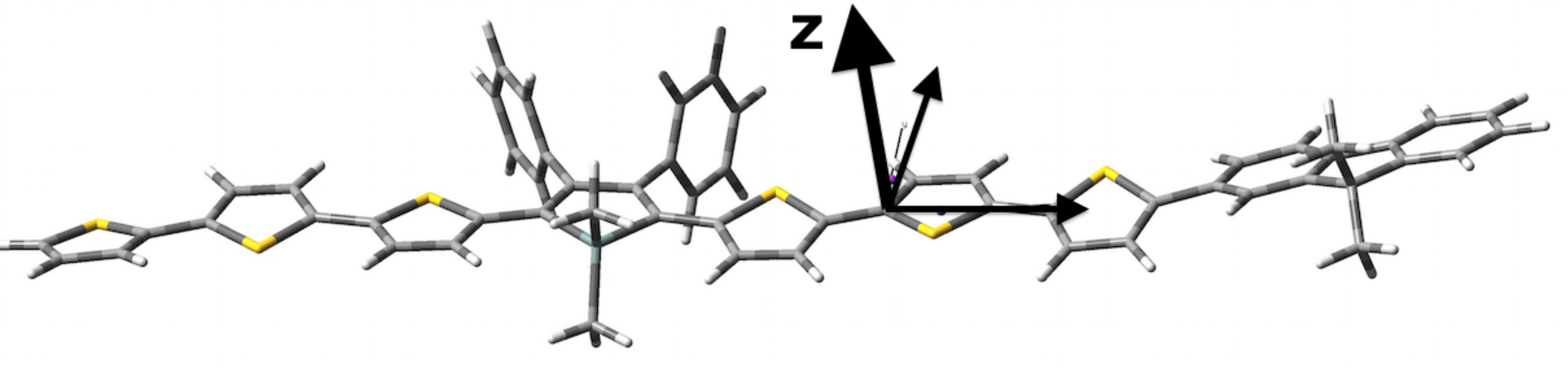}  
\end{center}
\caption{\label{REFE} Selected systems with their corresponding frame of reference:  {\bf a)} S8, {\bf b)} S31, {\bf c) S3},  {\bf d) }S27, {\bf e)} S47 } 
\end{figure}

The subset of molecules that present a planar geometry are systems: S1$>$S25$>$S13$>$S26$>$S33$>$S21 (from the highest to the lowest-energy values). This group is made of the fragments F3, F8, F12, F15, F18, F25, F26 or F27. In Fig. \ref{planarsys}, we present the planar systems, where no deviations were found at the corresponding building blocks (fragments). It may also be noticed that the 3D structures shown in Fig. \ref{planarsys}, present the corresponding saturations with H atoms to complete the corresponding valences. 
Such notation was omitted for clarity in the list of schematic representations presented in Fig. {\bf S.1-S.3} of SI. The remaining 44 molecular systems are non-planar structures that present a corresponding angle of torsion with respect to a given frame of reference as shown in Fig.\ref{REFE} for a group of 5 selected systems. 
The rest of the non-planar structures are reported in Fig. {\bf S.4-S.10} of SI. Particularly, we have also presented Table {\bf S.3-S.8} of SI  with a list of the 44 non-planar structures with respect to the angle of rotation observed in the respective component fragment. Such angle of rotation has been measured with regard to a specific axis, as shown in the systems of Fig. \ref{REFE}. The rest of the non-planar systems are all depicted in Fig. {\bf S.4-S.10} of SI. This group of systems are mainly composed by the F1, F3, F5, F22, and F25 fragments. The angles of rotation destabilize the total energy of systems up to 214 a.u., ranging from 70$^{\circ}$ to 105 $^{\circ}$ with respect to the specific axis shown in Fig. \ref{REFE}. This behavior was specifically found for systems S9, S10, S11, S13, S14, S15, S25, S26, S29, S36, S41, S43 and S44.

\section*{\sffamily \Large HOMO-LUMO gap}

After optimizing all molecular structures, the HOMO-LUMO gap was estimated. HOMO refers the Highest Occupied Molecular Orbital and LUMO corresponds to the Lowest Unoccupied Molecular Orbital. Such energetic levels may be interpreted as the valence band (HOMO) and the conduction band (LUMO), in terms of band theory. The energetic difference between these two levels may readily be interpreted as the band gap energy, which is a representative signature found on photovoltaic materials. It is well known that functional photovoltaic devices operate in a range from about 0.7 eV to 2.5 eV. In the systems under study, we found that the HOMO-LUMO gap range in the interval 1.15 eV to 3.0 eV. Exceptionally, systems S27 and S41 present a band gap of 5.019 and 7.397 eV, respectively. The systems under study, present a band gap in the following decreasing order of energy: S41 $>$ S27 $>$ S29 $>$ S9 $>$ S15 $>$ S2 $>$ S4 $>$ S31 $>$ S32 $>$ S30 $>$ S45 $>$ S14 $>$ S1 $>$ S26 $>$ S39 $>$ S36 $>$ S37 $>$ S42 $>$ S40 $>$ S25 $>$ S19 $>$ S43 $>$ S11 $>$ S10 $>$ S18$>$  S16 $>$  S03 $>$ S20 $>$ S44 $>$ S49 $>$ S50 $>$ S48 $>$ S46 $>$ S5 $>$ S38 $>$ S22 $>$ S12 $>$ S13 $>$ S7 $>$ S17 $>$ S47 $>$ S8 $>$ S6 $>$ S34 $>$ S35 $>$ S33 $>$ S23 $>$ S28 $>$ S24 $>$ S21.

\begin{figure}
\begin{center}
\includegraphics[width=0.6\columnwidth,keepaspectratio=true,angle=-90]{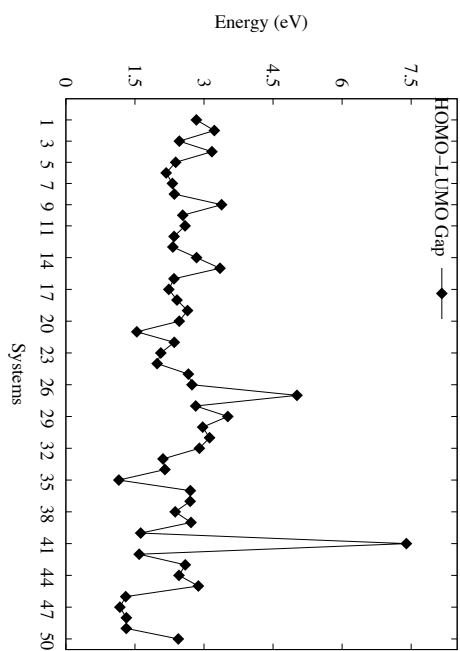}
\caption{\label{grafiGap}HOMO-LUMO gap energy of the 50 molecular systems.} 
\end{center}
\end{figure}

Besides, a schematic representation of the HOMO-LUMO gap is depicted in Fig. \ref{grafiGap}, and it may be verified that all energy gaps are in good agreement with the condition to be used as a PV material (i.e. with a band gap of about 2.0 eV). The failure to find an acceptable band gap on systems S27 and S41 (see Fig. \ref{grafiGap}) may be addressed to the omission of functional groups at the peripheral of such complexes or it may be directly ascribed to an effect that rises if more units of the polymer are also considered in the calculation. On the other hand, Scharber et al. \cite{Muhlbacher06} indicates that the energy conversion efficiency on an OPV may be estimated as a function of the band gap and the LUMO level of the donor. It was found that the energy conversion efficiency may be higher than 10\% for an energy gap smaller than 1.74 eV and a E$_{LUMO}$ $<$ -3.92 eV. Furthermore, an efficiency around 6\% may also be found for an $E_{gap}$ $<$ 2.15 eV and E$_{LUMO}$ $<$ -3.6 eV. Consequently, we presented the E$_{HOMO}$; E$_{LUMO}$ and E$_{gap}$ of the series of systems under study (see Table {\bf S.11-S.16}) and we assessed such data to find those OPVs that fit the criteria. It was found that such systems, according to decreasing order in energy gap are S34$>$S35$>$S33$>$S23$>$S28$>$S24$>$S21. Such systems may have high conversion energy efficiencies ranging from 6-10\%.

\section*{\sffamily \Large Open circuit voltage (V$_{OC}$)}

\begin{figure}
\begin{center}
\includegraphics[width=0.5\columnwidth,keepaspectratio=true ]{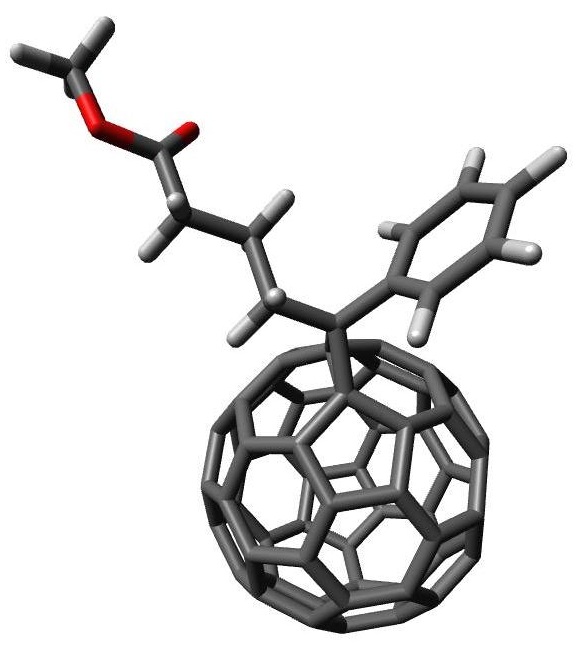}
\end{center}
\caption{\label{PCBM}Structural representation of  [6,6]-Phenyl-C61-butyric acid methyl ester (PCBM).}
\end{figure}

\begin{figure}
\begin{center}
\includegraphics[trim= 0.8in 1in 0.5in 1in,width=0.5\columnwidth,keepaspectratio=true,angle=-90]{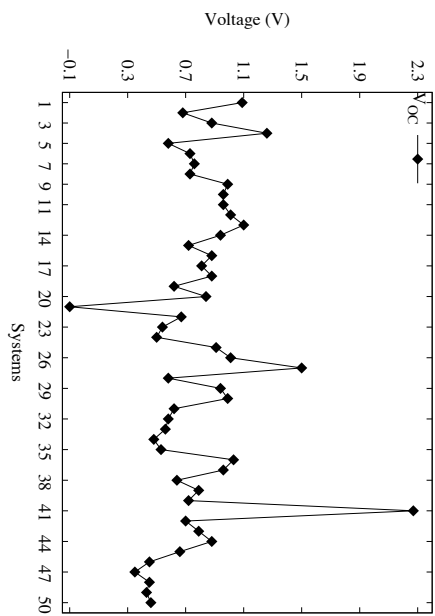}
\end{center}
\caption{\label{voc_figure}Open circuit voltage V$_{OC}$ of the 50 molecular systems under study.}
\end{figure}

The V$_{OC}$ provides information on the performance of a solar cell device into operation and in accordance with Eq. \ref{ecu-Voc}, V$_{OC}$ approaches the efficiency of the excitation dissociation process in order to allow the charge carriers to be conducted to the electrodes. To evaluate this approach, it is explicitly necessary to consider the E$_{LUMO}$ of an acceptor. It is known that PCBM ([6,6]-Phenyl-C61-butyric acid methyl ester) \cite{Horie:12} is one of the most widely used acceptors. As stated in the Computational Details section, it was found that the E$_{LUMO}$ of this acceptor system (see Fig. \ref{PCBM}) is -4.026 eV. This result was used to produce Fig. \ref{voc_figure}, where the V$_{OC}$ of the 50 systems under study are presented. The explicit data are also presented in Table {\bf S.11-S.16} of SI. The experimental data available for those properties are also presented for comparison. A reasonable agreement was found with the theoretical results found in this work. All values of V$_{OC}$ range from 0.35V to 2.27V, except system S21 (-0.10V), which reports a HOMO enegy close to the LUMO of the acceptor. The values for V$_{OC}$ decrease in the following order: S41 $>$ S27 $>$ S4 $>$ S13 $>$ S1 $>$ S36 $>$ S12 $>$ S26 $>$ S9  $>$ S30 $>$ S10 $>$ S11 $>$ S37 $>$ S14 $>$ S29 $>$ S25 $>$ S3 $>$  S16 $>$  S18 $>$ S44 $>$ S20 $>$ S17 $>$ S39 $>$ S43 $>$ S7 $>$ S6  $>$ S8 $>$ S15 $>$ S40 $>$ S42 $>$ S2 $>$ S22 $>$ S45 $>$ S38 $>$ S19 $>$ S31 $>$ S5 $>$ S28 $>$ S32 $>$ S33 $>$ S23 $>$ S35 $>$ S24 $>$ S34 $>$ S50 $>$ S46 $>$ S48 $>$ S49 $>$ S47 $>$ S21. We found no correlation among the V$_{OC}$ values and the planarity of the systems, since system S41 (see Fig. \ref{voc_figure}) presents non-planar symmetry and the highest V$_{OC}$ value, while system S21 with planar geometry presents the lowest V$_{OC}$, but no clear correspondence was verified from these data. Nevertheless, the presence of the F5 fragment induces a decrease in V$_{OC}$, locating systems S46-S50 with the lowest V$_{OC}$ in the order S50$>$S46$>$S48$>$S49$>$S47. All V$_{OC}$ values found in the training set fall in the range of available experimental data\cite{Horie:12}.

\begin{figure}
\begin{center}
 \qquad {\bf a) } LUMO of PCBM
 \vspace{0.3cm}
 
 \includegraphics[trim= 0.3in 1in 0.5in 0.5in,width=2in,angle=90]{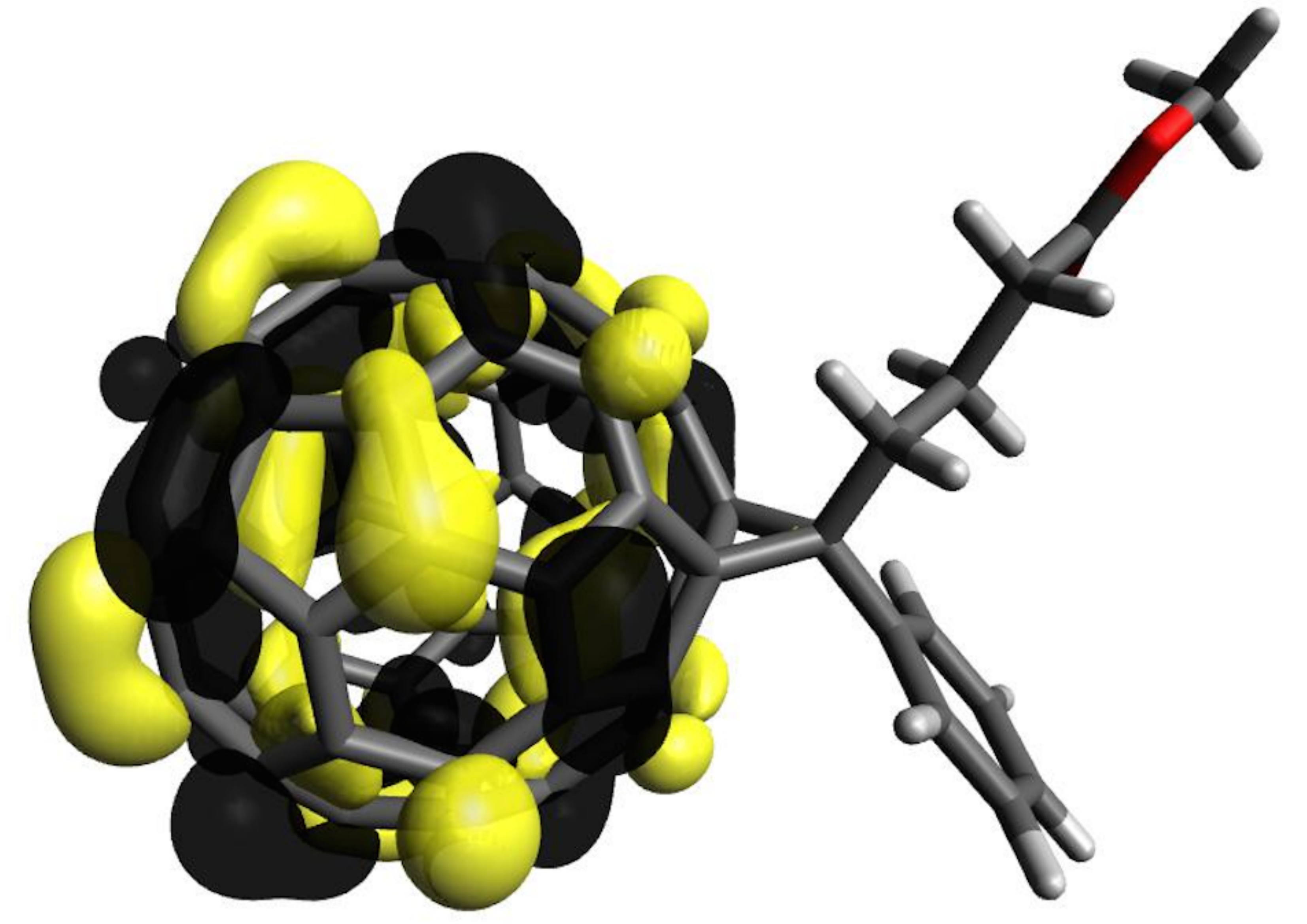}
\vspace{0.3cm}

 \includegraphics[trim= 0.3in 0.8in 0in 1.5in,width=2.6in]{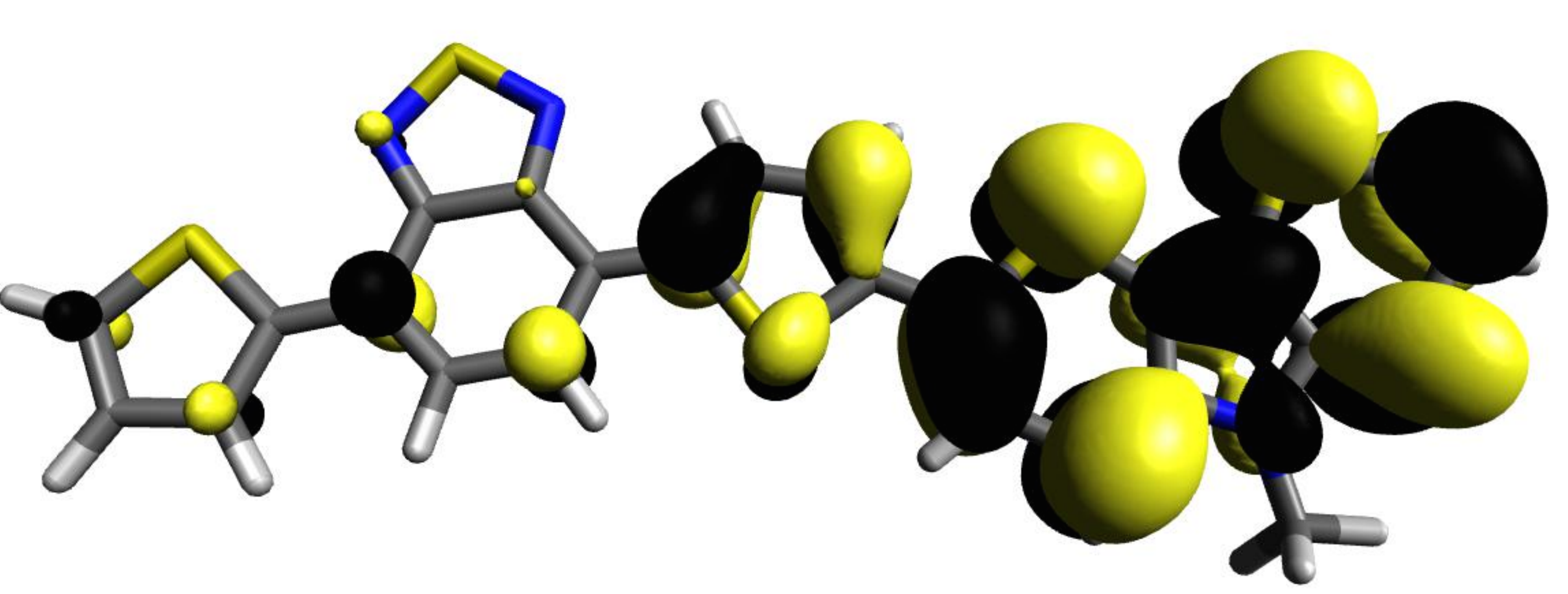} \quad \includegraphics[trim= 0.3in 0.8in 0in 0.7in,width=2.7in]{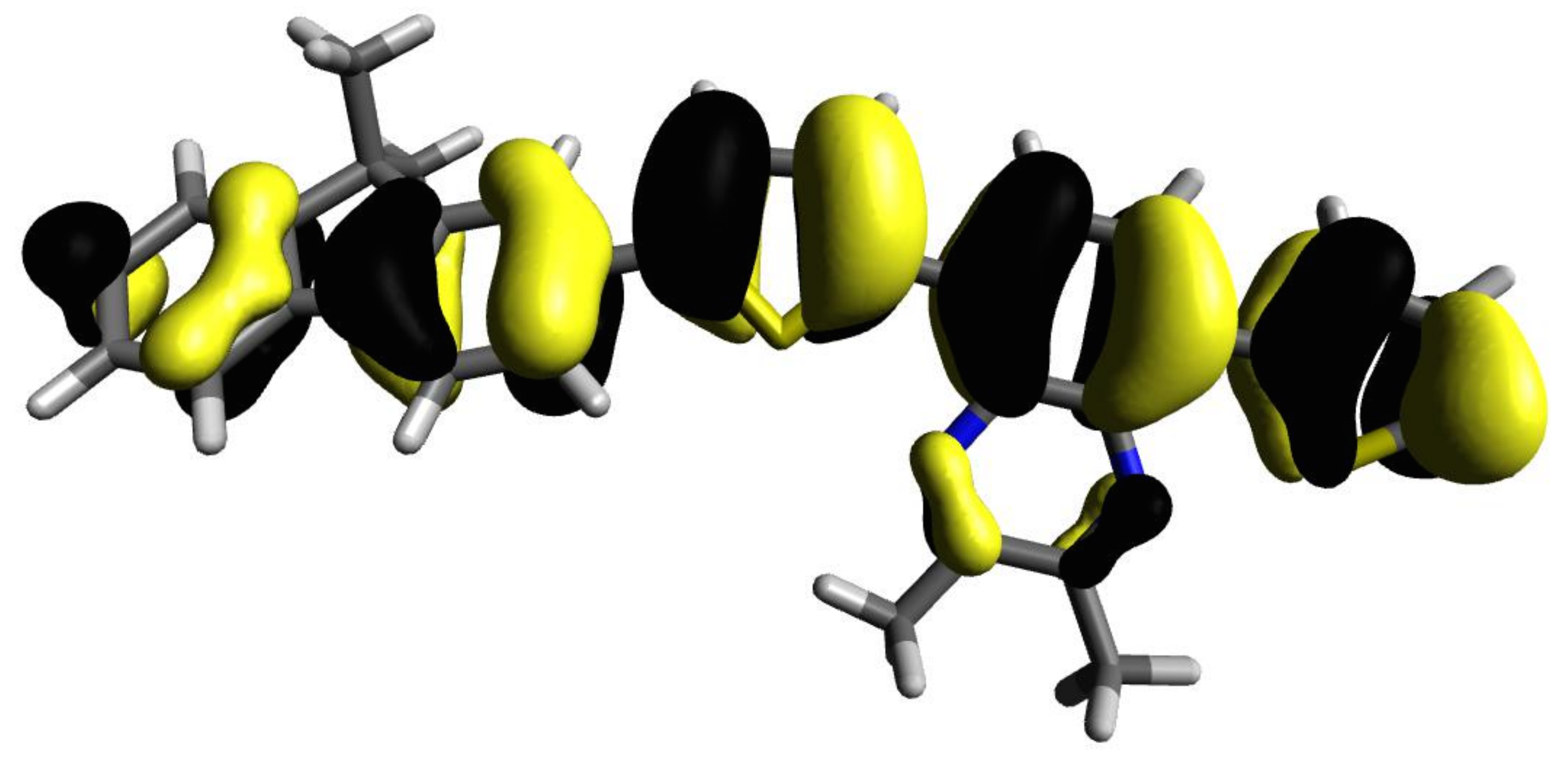}

\quad {\bf b ) } HOMO of S21 \qquad\qquad \qquad\qquad {\bf c) } HOMO of S41
\end{center}
\caption{ \label{voc_menor_mayor} {\bf a) } LUMO of PCBM; {\bf  b) } HOMO of S21; {\bf c) } HOMO of S41}
\end{figure}

Furthermore, the HOMO isosurfaces on systems S21 and S41, with the LUMO of the PCBM acceptor are shown in Fig. \ref{voc_menor_mayor}.  A distribution of $\pi$-bonding molecular orbitals are found for both systems, but an effective continuous distribution is found for the S41 system, and a distribution at random is obtained for the S21 system. The HOMO isosurfaces in those systems with highest V$_{OC}$ are depicted in Fig. {\bf S.11 } of SI and present that behavior; while the HOMO isosurfaces of the systems with smallest V$_{OC}$ are depicted in Fig. {\bf S.12} of SI. The $\pi$-bonding MOs present a discontinuous distribution through the covalent interaction between the carbon atoms. Consequently, an ordered distribution of $\pi$-orbitals along the chain of an OPV polymer in a solar cell, may be determinant to obtain a high output in V$_{OC}$ and may represent a criteria on the design of the device.

The PCBM LUMO isosurface presented in Fig. \ref{voc_menor_mayor}-{\bf (a)} represents a virtual orbital where no electronic charge resides, but it represents the probable regions around the PCBM where charge may be transferred to. In this case, it may be expected that the charge carriers jump to the C$_{60}$ unit of the PCBM, and unlikely to the attached functional groups. On the other hand, it has been stated\cite{Scharber06}  that if the difference between the LUMO of the donor and acceptor ($\Delta E_{LL}$) lies about 0.3 eV (i.e. LUMO$_D$ - LUMO$_A$ = 0.3 eV), a larger energy conversion efficiency is obtained. Consequently, we found that $\Delta E_{LL}$ for most of the systems under study ranges from 0.92 to 1.2 eV. The systems are ordered in the following decreasing order of $\Delta E_{LL}$: S41 $>$ S27 $>$ S15 $>$ S2 $>$ S29 $>$ S31 $>$ S32 $>$ S45 $>$ S19 $>$ S42 $>$ S49 $>$ S30 $>$ S39 $>$ S40 $>$ S46 $>$ S48 $>$ S50 $>$     S4 $>$ S14 $>$ S43 $>$ S47 $>$ S5 $>$ S1 $>$ S22 $>$ S25 $>$ S26 $>$ S37 $>$ S38 $>$ S20 $>$ S21 $>$ S34 $>$ S35 $>$ S36 $>$ S44 $>$ S3 $>$ S16 $>$ S18     $>$ S9 $>$ S10 $>$ S11 $>$ S7 $>$ S23 $>$ S24 $>$ S33 $>$ S6 $>$ S8 $>$ S28 $>$ S17 $>$ S12 $>$ S13. Using the criteria for V$_{OC}$ and $\Delta E_{LL}$, we found that the 15 systems closest to 0.3 eV are S16, S18, S9, S10, S11, S7, S23, S24, S33, S6, S8, S28, S17, S12, S13. If we consider the systems with largest V$_{OC}$ and smallest $\Delta E_{LL}$, the potential systems to be implemented as OPVs would be S13 $>$ S12 $>$ S17 $>$ S28 $>$ S8.

\section*{\sffamily \Large Aromaticity properties}

\begin{figure}
\begin{center}
\includegraphics[trim= 0.8in 1in 0.5in 1in,width=0.5\columnwidth,keepaspectratio=true,angle=-90]{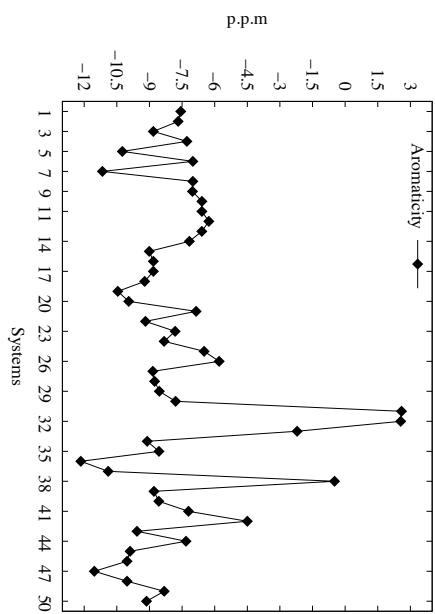}
\end{center}
\caption{\label{nics}  NICS values of aromaticity in the 50 molecular systems.}
\end{figure}

We performed NICS (Nuclear Independent Chemical Shifts) calculations on the 50 optimized systems according to the methodology described in the Computational Details Section in order to understand the possible aromatic behavior observed in the systems under study. We introduced the corresponding ghost atom at a location close to a geometrical center on each of the systems as it is indicated in Fig. {\bf S.13 - S.15 } for the 50 molecular systems. The value of the NICS indexes decrease in the following order along the series: S36 $>$ S47 $>$  S7  $>$ S37 $>$ S19 $>$ S5 $>$ S20 $>$ S45 $>$ S43 $>$ S22 $>$ S50 $>$ S34 $>$ S15 $>$ S27 $>$ S18 $>$ S16 $>$ S3 $>$ S39 $>$ S28 $>$ S48 $>$ S46 $>$ S40 $>$ S35 $>$ S29 $>$ S24 $>$ S49 $>$ S23 $>$ S30 $>$ S2 $>$ S1 $>$ S44 $>$ S4 $>$ S41 $>$ S14 $>$ S9 $>$ S8 $>$ S6 $>$ S17 $>$ S21 $>$ S13 $>$ S11 $>$ S10 $>$ S25 $>$ S12 $>$ S26 $>$ S42 $>$ S33 $>$ S38 $>$ S32 $>$ S31.  It was found a strong aromatic behavior in the majority of the complexes (see Fig. \ref{nics}), and a NICS value greater than -7.0 ppm (corresponding to benzene) was found. This indicates an enhanced aromatic character through the series, which is only altered for systems S31, S32 and S38 that present an antiaromatic behavior. This may be attributed to the location of the ghost atom, which remains in the center of a ring connected to 3 or more groups.

The presence of such aromatic groups may rearrange the electronic distribution of the $\pi$-orbitals and destabilize the aromatic character at that location. Furthermore, according to Mu\~niz et al., the aromatic character of a system is correlated with the conductivity \cite{Muniz13}, since charge transfer may be achieved through the isotropic currents at the center of the aromatic rings in a polymerized arrangement, such as the one-dimensional chain of the Au$_3$Cl$_3$Li$_2$ system \cite{Muniz13}. In the OPV system under study, this property may be relevant since the synthesis of such organic compounds may be designed in columnar arrays, where the electronic transport may be facilitated. In agreement with the latter, the potential candidates to be polymerized in columnar arrays may correspond to those cases where the aromatic character is stronger, i.e, for the systems S36 $>$ S47 $>$  S7  $>$ S37 $>$ S19 $>$ S5 $>$ S20 $>$ S45 $>$ S43 $>$ S22 $>$ S50 $>$ S34 $>$ S15 $>$ S27 $>$ S18.The electron-hole pair in these systems may provide a more effective charge transport to an acceptor in the design of a solar cell.

\section*{\sffamily \Large Non-linear optical (NLO) properties}

\begin{figure}
\begin{center}
\includegraphics[trim= 0.8in 1in 0.5in 1in,width=0.5\columnwidth,keepaspectratio=true,angle=-90]{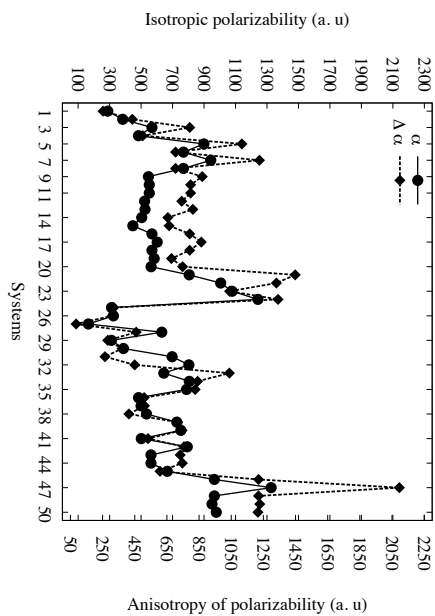}
\end{center}
\caption{\label{NLO} Isotropic polarizability $\alpha$ and anisotropy of polarizability $\Delta \alpha$ values of the 50 molecular systems.}
\end{figure}

\begin{figure}
\begin{center}
\includegraphics[trim= 0.8in 1in 0.5in 1in,width=0.5\columnwidth,keepaspectratio=true,angle=-90]{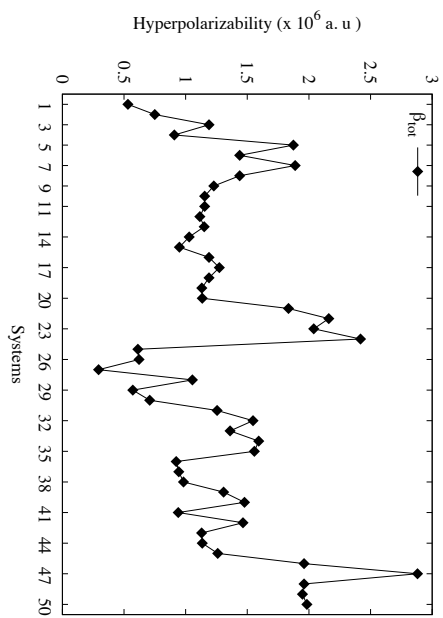}
\end{center}
\caption{\label{NLO_beta} Hyperpolarizability $\beta_{tot}$ values of the 50 molecular systems.}
\end{figure}

NLO properties, such as the isotropic polarizability ($\alpha$), anisotropy of polarizability ($\Delta\alpha$) and total hyperpolarizability ($\beta_{tot}$), are of high relevance to understand the behavior and performance of OPVs. NLO properties are also of interest in materials for modern communication technologies and optical signals processing\cite{Asiri20}. NLO properties give a tendency on the delocalization of intramolecular charge in groups of donor electrons\cite{Senge07}. Further, NLO properties relate higher efficiency of mobility and electron transport from donor to acceptor to higher $\alpha$, $\Delta\alpha$ and $\beta_{tot}$. We computed such properties for the 50 systems under study. The values for $\alpha$ and $\Delta\alpha$ are depicted in Fig. \ref{NLO}, while those for $\beta_{tot}$ are presented in Fig. \ref{NLO_beta}. All parameters are also reported in  Table {\bf S.9} of of SI. After selecting those systems with the highest values for the NLO properties, we found that systems S5, S7, S21, S22, S23, S24, S46, S47, S49 and S50 would represent potential systems that would effectively transfer electronic charge from the donor to the acceptor.

\section*{\sffamily \Large Excited state calculations}

\subsection*{\sffamily \large UV/vis spectra}

\begin{figure}
\begin{center}
\includegraphics[trim= 0.8in 1in 0.5in 1in,width=0.5\columnwidth,keepaspectratio=true,angle=-90]{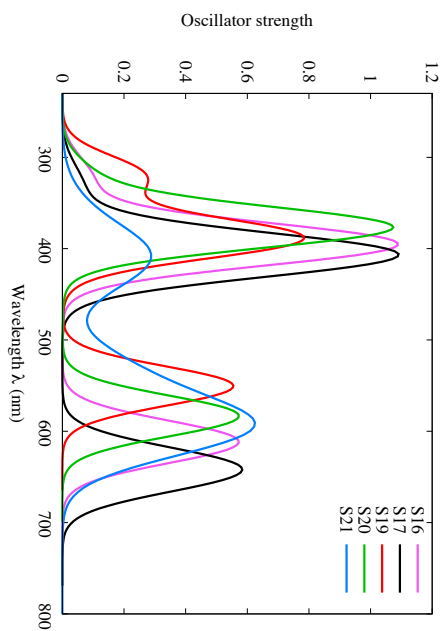}
\end{center}
\caption{\label{S16_S21} UV/vis spectra of S16, S17, S19, S20, S21.}
\end{figure}

\begin{figure}
\begin{center}
\includegraphics[trim= 0.8in 1in 0.5in 1in,width=0.5\columnwidth,keepaspectratio=true,angle=-90]{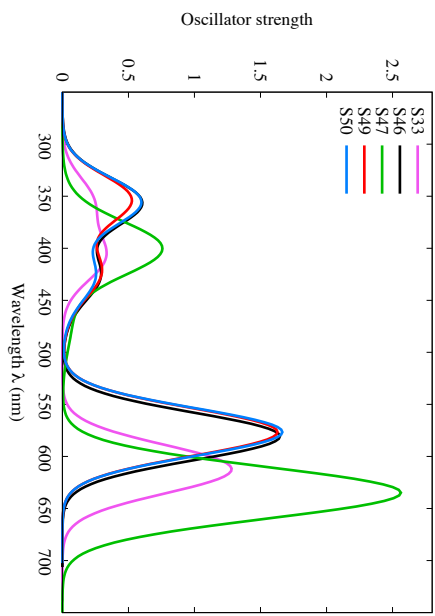}
\end{center}
\caption{\label{S33_S50} UV/vis spectra of S33, S46, S47, S49, S50.}
\end{figure}

Excited state calculations were carried out on a series of candidate OPV systems using the TD-DFT methodology as presented in the Computational Details section. Such calculations were performed by considering the ground state geometry of the 50 systems with the first 10 excited states. The results for the lowest-energy excited states are presented in Table \ref{estado_exitado}. The locations of the adsorption bands for a group of selected systems and their corresponding oscillator strengths reveal that all systems present two adsorption bands separated by a corresponding Stokes shift. Note that the oscillator strength magnitude may be interpreted as a measurement of the probability to find an absorption peak at a corresponding wavelength. Furthermore, the simulated UV/vis spectra were also computed from the excited state calculations (see Fig. \ref{S16_S21} and \ref{S33_S50}). All systems presented in Table \ref{estado_exitado} were selected from the rest of the calculations since all absorption bands are located in the range of visible light (from about 350 to 700 nm) and all systems are summarized in Table {\bf S.17-S.21} of SI. This makes the selected systems suitable to be implemented in applications of solar cell materials. We further used Gouterman’s 4-Molecular Orbital (MO) model \cite{Gouterman:61} to analyze the individual transitions, which was originally considered to describe the excitations of porphyrins. The MOs to be considered are HOMO-1, HOMO, LUMO and LUMO+1. The model states that the transitions may take the following excitations: HOMO-1 $\rightarrow$ LUMO+1,  HOMO $\rightarrow$ LUMO+1, HOMO-1$\rightarrow$ LUMO and HOMO $\rightarrow$ LUMO, which are designated as $B_y$, $B_x$, $Q_x$ and $Q_y$ bands. The molecular systems S16 to S21 present bands with higher oscillator strengths located at about 400 nm, and a band with smaller oscillator strength located around 600 nm. On the other hand, the absorption bands for systems S33, S46-S50 present bands with a small intensity and negligible oscillator strength. As it may be readily verified in Table \ref{estado_exitado}, the transitions for the selected systems comprise to a B and a Q band, namely $B_x$ and $Q_y$ bands. The $Q_y$-bands present the highest probability of occurrence. Such behavior is virtually found along the complete series (see Table {\bf S.17-S.21} of SI), with an exception at the transition HOMO $\rightarrow$ LUMO+2, which was found for systems S38, S39 and S41, corresponding to transitions with a low probability of occurrence. The highest oscillator strengths correspond to those systems that present red-shifting, and their corresponding absorption bands are located from 550 to 650nm, which indicates that such systems would absorb most of the visible light. The UV/vis spectra for the rest of the systems under study may be found from Fig. {\bf S.16 } to {\bf S.25} of SI. In order to elucidate the changes in the electronic structure that induce the corresponding excited states, we considered the $|CI|$-coefficients that provide information on the probability of occurrence of a given transition. We present those excitations with the largest $|CI|$-coefficients for the selected systems presented in Table \ref{estado_exitado}.

\begin{figure}
\begin{center}
{\bf a)}  S17 \qquad \qquad\qquad\qquad\qquad\qquad \qquad {\bf b)}  S47    \qquad \qquad  \\
\vspace{0.2cm}

\includegraphics[width=3in]{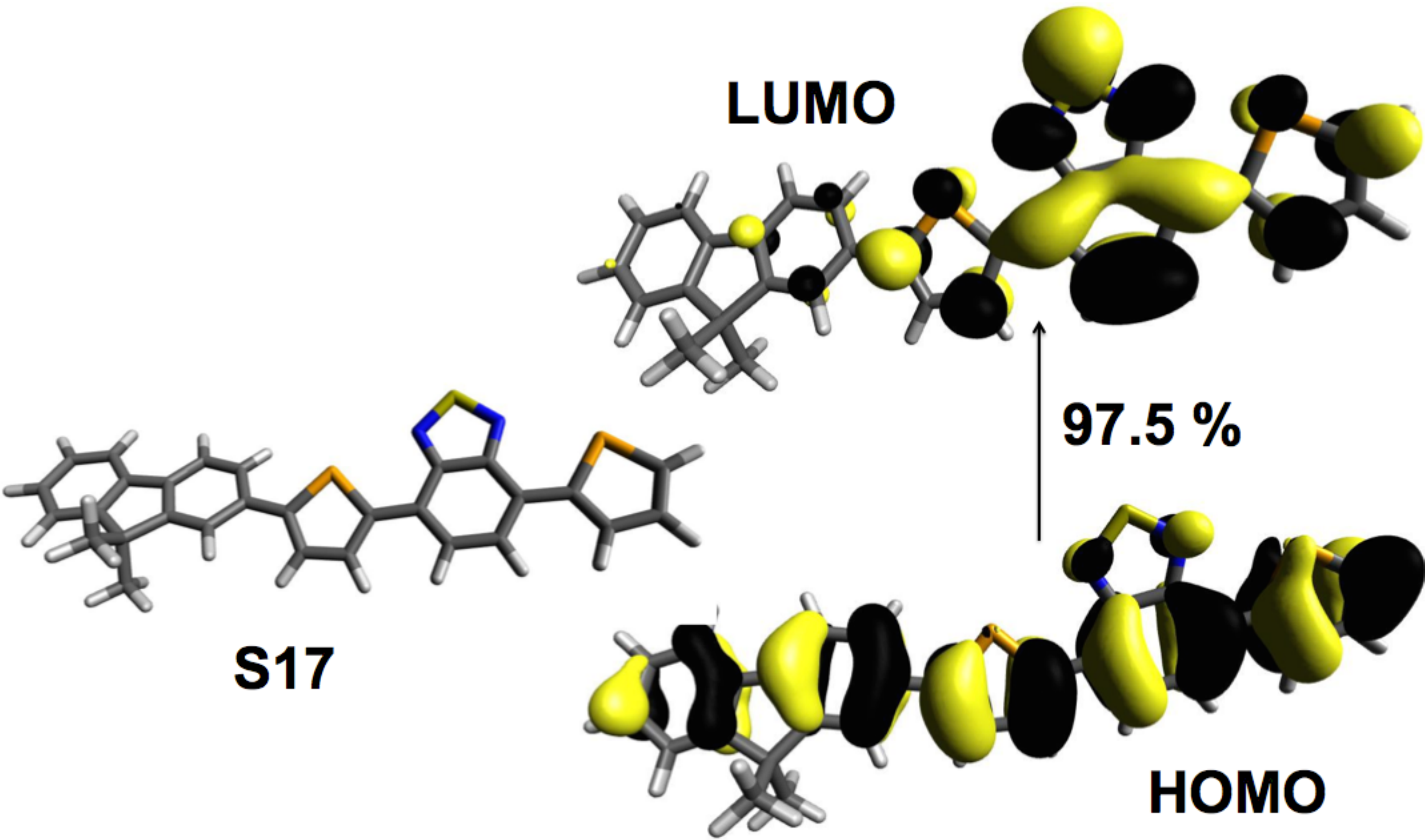} \includegraphics[width=3in]{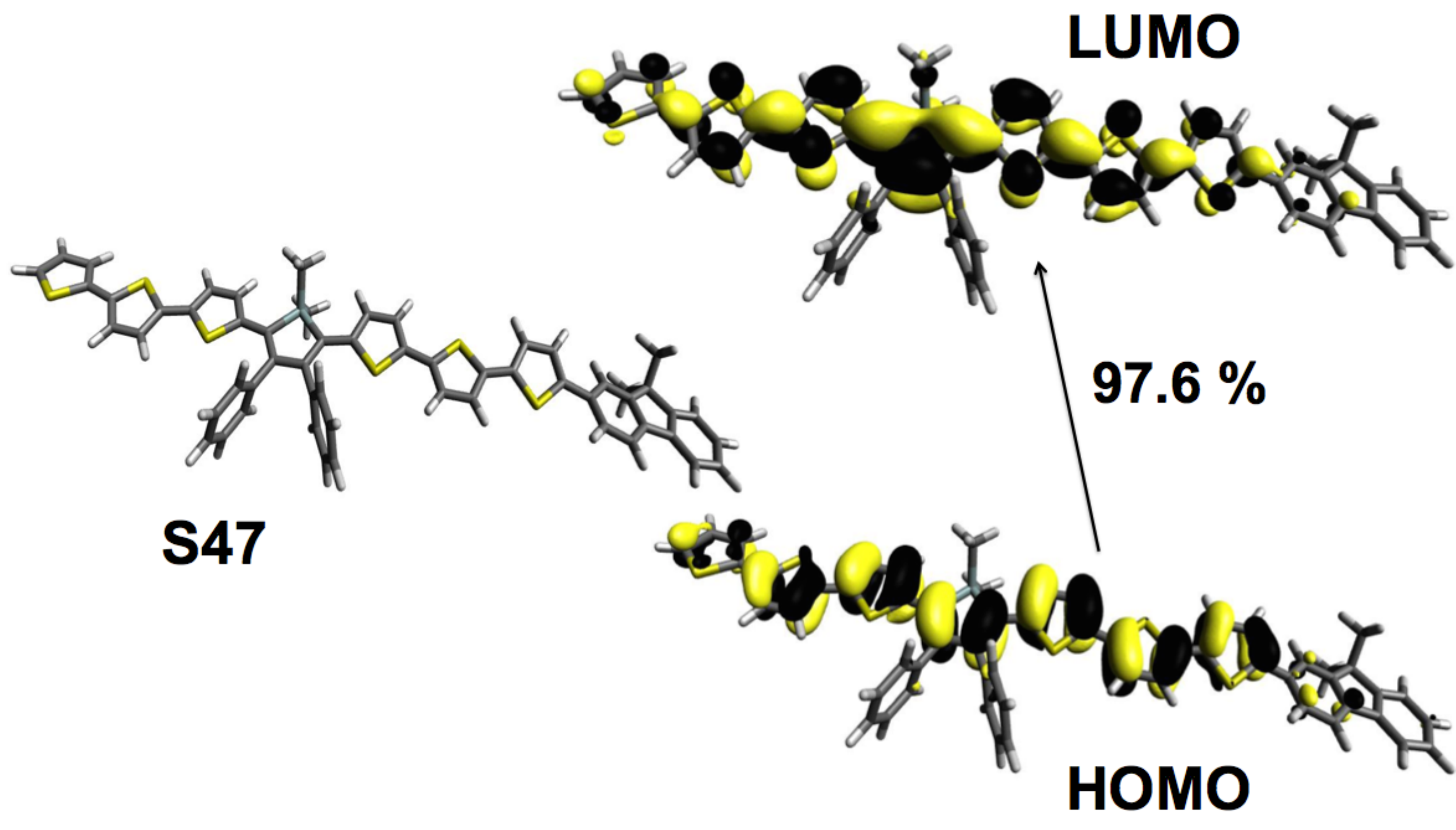}
\vspace{0.2cm}

{\bf c) }  S50
\vspace{0.2cm}

\includegraphics[width=3in]{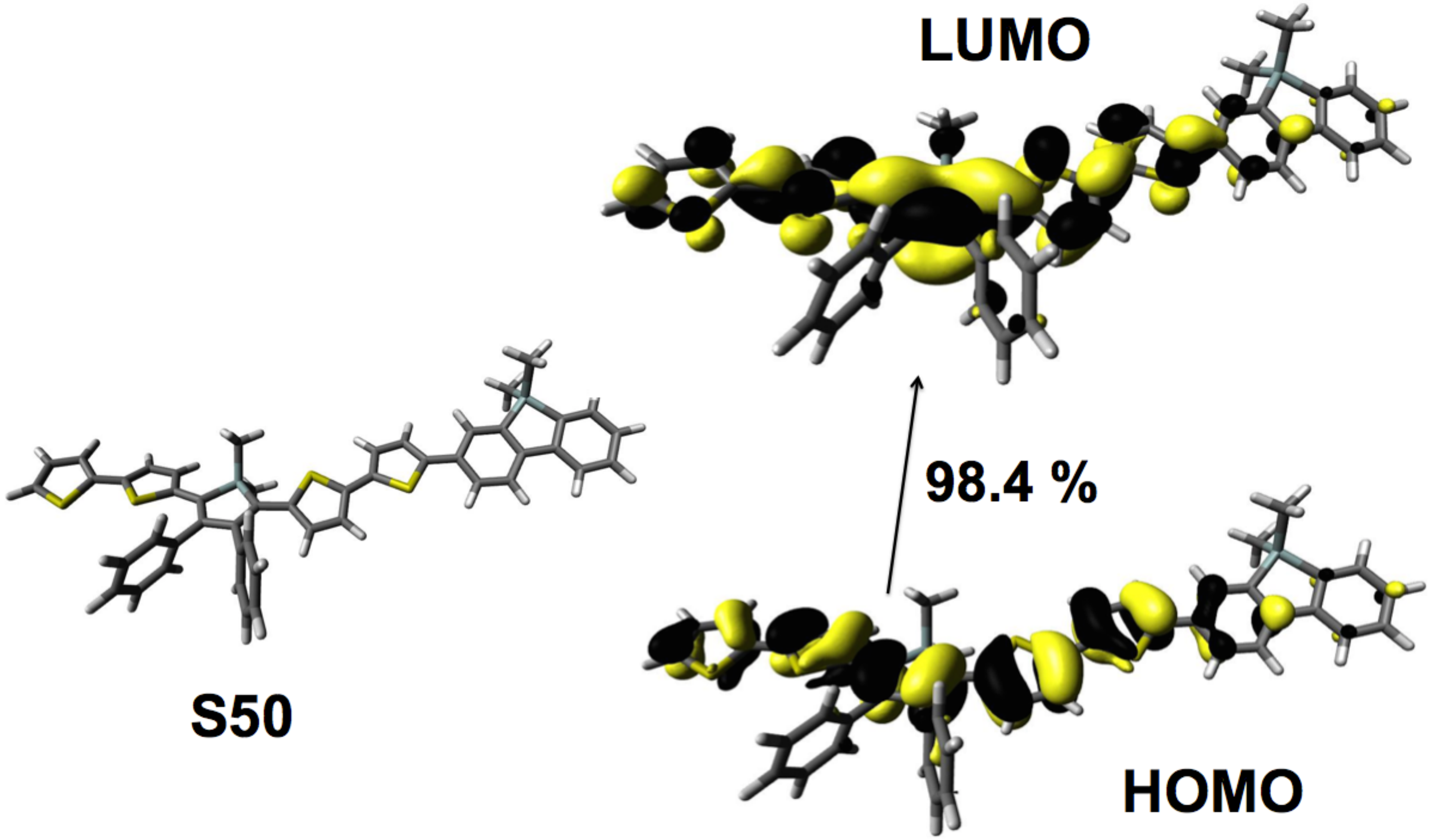}
\end{center}

\caption{\label{Transition}Isosurfaces for electronic transitions: {\bf a)} S17, {\bf b)} S47, {\bf c) } S50}
\end{figure}

Such coefficients provide information on the probability of occurrence of the given transition. We present those excitations with the largest  coefficient for the selected systems presented in Table  \ref{estado_exitado}. A general behavior may be verified in this series of systems. That is, all transitions are more likely to occur (see Fig. \ref{Transition}) from the HOMO to the LUMO of the first three lowest-energy systems with a probability ranging from 97.5\% to 98.5\%. Note that the rest of excitations are depicted in Fig. {\bf S.26-S.34} of SI. The one-electron transition takes place from the HOMO composed of $\pi$-orbitals mainly located at the carbon atoms to the virtual molecular orbital LUMO, which is mainly composed of anti bonding $\pi^{*}$ orbitals. Note that the LUMO represents the most likely regions on the molecular arrangement where electronic charge may be transferred to, since it represents an unoccupied MO. The electronic transfer may be ascribed to an intraligand (IL) transition with a strong $\pi-\pi^{*}$ character. Such charge transfer is directly related to a low-energy absorption that corresponds to a small HOMO-LUMO gap as presented in Table {\bf S.11-S.16} of SI  and Fig.\ref{grafiGap}. The LUMO may be interpreted as the hole that the electron must track in order to achieve an electron-hole pair process, describing the mechanism of charge transport in the photovoltaic material. Since the $Q_y$-bands are the most likely excitations of the series, and they correspond to the red-shifted bands that would absorb a wider range of the visible light, it may be expected that the series of title complexes would all effectively absorb light in photovoltaic applications. The HOMO-LUMO gap for this series of systems range from 1.15 to 3.0 eV, which corresponds to the band gap of known photovoltaic semiconductors such as a-Si\cite{Wronski:04} or AlGaAs\cite{Soga:99}. Consequently, the selected series of systems would represent potential candidates to be implemented as an alternative to the based-inorganic PV materials.

\subsection*{\sffamily \large Light Harvesting Efficiency calculations}

\begin{figure}
\begin{center}
\includegraphics[trim= 0.8in 1in 0.5in 1in,width=0.5\columnwidth,keepaspectratio=true,angle=-90]{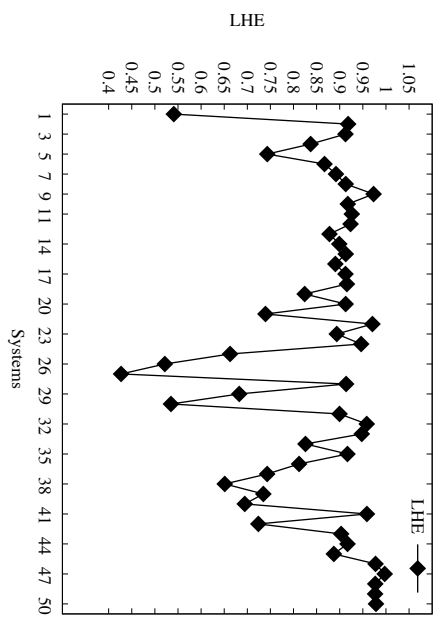}
\end{center}
\caption{\label{LHE_graph}  Light Harvesting Efficiency (LHE) values of the 50 molecular systems.}
\end{figure}

As we have previously reported, all the excited states for the 50 systems in the training set are listed in Table  \ref{estado_exitado}. The oscillator strengths ($f$) for all systems are also reported in this table and it provides information on the intensity of the absorption band, as calculated for the corresponding excited state. As it has been previously stated in the Computational Details section, the Light Harvesting Efficiency (LHE) may be approached by Eq. \ref{LHE}, indicating that the LHE is proportional to the oscillator strength intensity. On the other hand, LHE values are correlated with the intensity of the short circuit current density ($J_{SC}$). Fig. \ref{LHE_graph} shows the behavior of LHE through the set of systems, indicating that those with larger LHE are S47 $>$ S50 $>$ S46 $>$ S49 $>$ S33 $>$ S17 $>$ S16 $>$ S20 $>$ S19 $>$ S21. These systems may be considered as potential OPVs that would increase the  $J_{SC}$ in a PV device. It was also found that this group of systems present a non-planar geometry with torsion angles from 30$^{\circ}$  to 60$^{\circ}$. In this set, the presence of Sulfur and Silicon in fragments F3 and F5 remains constant and may be directly responsible for the rotations around the corresponding axis of symmetry. Consequently, the implementation of non-planar OPVs in solar cell devices may increase density currents, which may be desirable in the design of solar cell materials.

\section*{\sffamily \Large Electron injection efficiency calculations}

\begin{figure}
\begin{center}
\includegraphics[trim= 0.8in 1in 0.5in 1in,width=0.5\columnwidth,keepaspectratio=true,angle=-90]{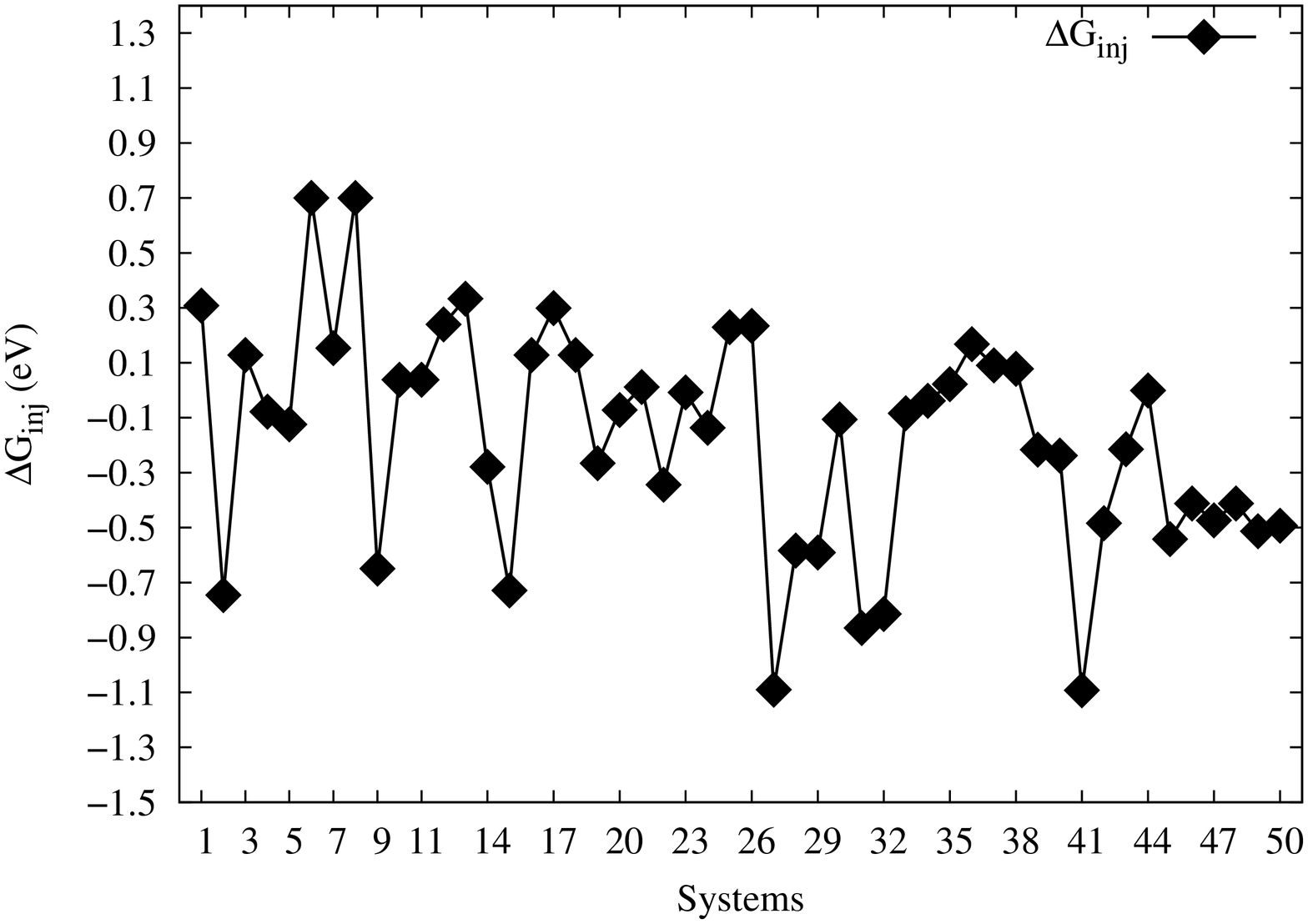}
\end{center}
\caption{\label{delta_G} $\Delta G_{inj}$ values for the 50 molecular systems.}
\end{figure}

The electron injection efficiency $\phi_{inj}$ may be approached by $\Delta G_{inj}$, as it is given in Eq. \ref{delta_inye}. This parameter is also related with the intensity of $J_{SC}$. Hence, for high values of LHE and $\Delta G_{inj}$, we may obtain enhanced values of $J_{SC}$. Using the available data for the oxidation potentials reported in Table {\bf S.10} of SI; the energies of the first excited state transitions found at the TD-DFT level and the LUMO of the PCBM system, as the acceptor, it was found all $\Delta G_{inj}$ values, as presented in Fig. \ref{delta_G}. $\Delta G_{inj}$ values were found in the following decreasing order of energy: S6 $>$ S8 $>$ S13 $>$ S1 $>$ S17 $>$ S12 $>$ S26 $>$ S25 $>$ S36 $>$ S7 $>$ S3 $>$ S16 $>$ S18 $>$ S37 $>$ S38 $>$ S10 $>$ S11 $>$ S35 $>$ S21 $>$ S44 $>$ S23 $>$ S34 $>$ S20 $>$ S4 $>$ S33 $>$ S30 $>$ S5 $>$ S24 $>$ S43 $>$ S39 $>$ S40 $>$ S19 $>$ S14 $>$ S22 $>$ S46 $>$ S48 $>$ S47 $>$ S42 $>$ S50 $>$ S49 $>$ S45 $>$ S28 $>$ S29 $>$ S9 $>$ S15 $>$ S2 $>$ S32 $>$ S31 $>$ S27 $>$ S41. Besides, if negative values for  $\Delta G_{inj}$ are found, an spontaneous electron injection to the LUMO of the acceptor may be addressed \cite{Yang12}. This was found for systems S2, S4, S5, S9, S14, S15, S19, S20, S22, S23, S24, S27-S34, S39-S50. The systems with $\Delta G_{inj}$ above 0.6 eV are of special interest \cite{Ding13, Yang12, Zhang13} to obtain enhanced $J_{SC}$ values. In our calculations, those values correspond to systems S6 and S8.

\begin{figure}
\begin{center}
\includegraphics[trim= 0.8in 1in 0.5in 1in,width=0.5\columnwidth,keepaspectratio=true,angle=-90]{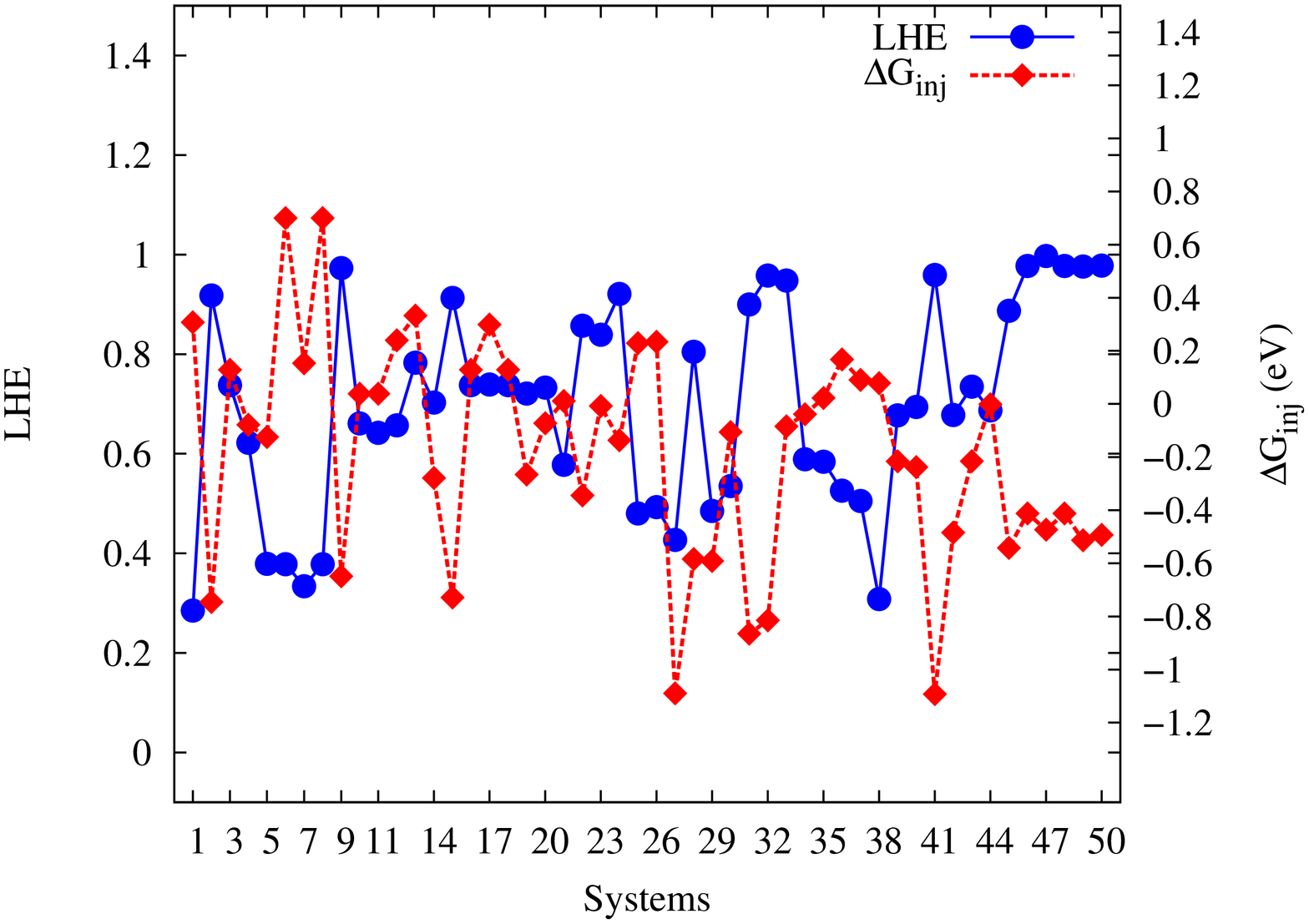}
\end{center}
\caption{\label{LHE_delta_G} LHE and $\Delta G_{inj}$ values for the 50 molecular systems.}
\end{figure}

Since $J_{SC}$ depends on LHE and $\Delta G_{inj}$, we have plotted both properties in Fig. \ref{LHE_delta_G} in order to relate the highest LHE with the highest $\Delta G_{inj}$ values. The combined properties would represent a final estimation on the tendency for $J_{SC}$ values found for this training set. Taking into account the former criteria, we found that the final group of systems with enhanced $J_{SC}$ values would be formed by the S6, S8, S13, S16, S17, S19 and S28 OPVs.

\section*{\sffamily \Large Conclusions}

An electronic structure investigation was performed on a series of OPV materials. The systems under study are divided into a group of planar structures and into structures with torsional angles about a symmetry axis along the molecules. The non-planar OPVs appear to be the most energetically stable systems in the series, i.e. systems S17, S22, S23, S24, S47, S48 and S50. The HOMO energy calculations show a reasonable agreement of 90 to 95\% with respect to experiment, and a 75 to 86\% of agreement in the LUMO energy calculations. This may be highly relevant in the prediction of photovoltaic properties. The aromaticity calculations reveal a high stability through the series, and particularly on systems S5, S7, S19, S20, S22, S36, S37, S43, S45 and S47. This property may be directly applied in the polymerization of the OPV material, where the molecules may be arranged in columnar phases that interact through $\pi-\pi$ stacking along the aromatic rings. This may result in a solid-state material with desirable electronic conductivity properties. On the other hand, the computed UV/vis spectra show that the most probable excitations would correspond to red-shifted $Q_y$-bands, that would absorb visible light more effectively. This may be addressed to an adequate performance in solar cell device applications.
The performance of this set of complexes was also characterized with the evaluation of LHE, $\Delta E_{LL}$ and $\Delta G_{inj}$, which were computed in the presence of a PCBM acceptor. After selecting the systems with highest NLO parameters and larger V$_{OC}$ values, it was found that the systems S6, S10, S12, S13, S16, S17, S23 and S33 would represent a set of optimal materials to be implemented in solar cell devices. After following the methodology developed in the present work, it may be possible to improve the design of solar cell devices based upon OPV structures.


\subsection*{\sffamily \large Acknowledgements}


J.M. wants to acknowledge the support given by C\'atedras-CONACYT (Consejo Nacional de Ciencia y Tecnolog\'ia) under Project No. 1191; the support given by CONACYT through Project SEP-Ciencia B\'asica No.156591; DGTIC (Direcci\'on General de C\'omputo y de Tecnolog\'ias de Informaci\'on y Comunicaci\'on) and the Supercomputing Department of Universidad Nacional Aut\'onoma de M\'exico for the computing resources under Project No. SC16-1-IR-29. A.G.L. wants to acknowledge the MSc. Scholarship provided by CONACYT with No. 376934 and PhD scholarship No. 306891. The authors thank Laboratorio de Supercómputo de Alto Rendimiento from Universidad Politécnica de Chiapas for the computing resources.


\clearpage

\bibliography{gold_gen}   

\begin{thebibliography}{51}
\expandafter\ifx\csname natexlab\endcsname\relax\def\natexlab#1{#1}\fi
\expandafter\ifx\csname bibnamefont\endcsname\relax
  \def\bibnamefont#1{#1}\fi
\expandafter\ifx\csname bibfnamefont\endcsname\relax
  \def\bibfnamefont#1{#1}\fi
\expandafter\ifx\csname citenamefont\endcsname\relax
  \def\citenamefont#1{#1}\fi
\expandafter\ifx\csname url\endcsname\relax
  \def\url#1{\texttt{#1}}\fi
\expandafter\ifx\csname urlprefix\endcsname\relax\def\urlprefix{URL }\fi
\providecommand{\bibinfo}[2]{#2}
\providecommand{\eprint}[2][]{\url{#2}}

\bibitem[{\citenamefont{Chaar et~al.}(2011)\citenamefont{Chaar, Lamont, and
  Zein}}]{Chaar11}
\bibinfo{author}{\bibfnamefont{L.~E.} \bibnamefont{Chaar}},
  \bibinfo{author}{\bibfnamefont{L.}~\bibnamefont{Lamont}}, \bibnamefont{and}
  \bibinfo{author}{\bibfnamefont{N.~E.} \bibnamefont{Zein}},
  \bibinfo{journal}{Renew. Sust. Energ. Rev.} \textbf{\bibinfo{volume}{15}},
  \bibinfo{pages}{2165} (\bibinfo{year}{2011}).

\bibitem[{\citenamefont{de~Wild et~al.}(2011)\citenamefont{de~Wild, Meijerink,
  Rath, van Sark, and Schropp}}]{Wild11}
\bibinfo{author}{\bibfnamefont{J.}~\bibnamefont{de~Wild}},
  \bibinfo{author}{\bibfnamefont{A.}~\bibnamefont{Meijerink}},
  \bibinfo{author}{\bibfnamefont{J.~K.} \bibnamefont{Rath}},
  \bibinfo{author}{\bibfnamefont{W.~G. J. H.~M.} \bibnamefont{van Sark}},
  \bibnamefont{and} \bibinfo{author}{\bibfnamefont{R.~E.~I.}
  \bibnamefont{Schropp}}, \bibinfo{journal}{Energy Environ. Sci.}
  \textbf{\bibinfo{volume}{4}}, \bibinfo{pages}{4835} (\bibinfo{year}{2011}).

\bibitem[{\citenamefont{Solangi et~al.}(2011)\citenamefont{Solangi, Islam,
  Saidur, Rahim, and Fayaz}}]{Solan11}
\bibinfo{author}{\bibfnamefont{K.}~\bibnamefont{Solangi}},
  \bibinfo{author}{\bibfnamefont{M.}~\bibnamefont{Islam}},
  \bibinfo{author}{\bibfnamefont{R.}~\bibnamefont{Saidur}},
  \bibinfo{author}{\bibfnamefont{N.}~\bibnamefont{Rahim}}, \bibnamefont{and}
  \bibinfo{author}{\bibfnamefont{H.}~\bibnamefont{Fayaz}},
  \bibinfo{journal}{Renew. Sust. Energ. Rev.} \textbf{\bibinfo{volume}{15}},
  \bibinfo{pages}{2149 } (\bibinfo{year}{2011}).

\bibitem[{\citenamefont{O'Regan and Gratzel}(1991)}]{ORegan91}
\bibinfo{author}{\bibfnamefont{B.}~\bibnamefont{O'Regan}} \bibnamefont{and}
  \bibinfo{author}{\bibfnamefont{M.}~\bibnamefont{Gratzel}},
  \bibinfo{journal}{Nature} \textbf{\bibinfo{volume}{353}},
  \bibinfo{pages}{737} (\bibinfo{year}{1991}).

\bibitem[{\citenamefont{Sayama et~al.}(2000)\citenamefont{Sayama, Hara, Mori,
  Satsuki, Suga, Tsukagoshi, Abe, Sugihara, and Arakawa}}]{Sayama00}
\bibinfo{author}{\bibfnamefont{K.}~\bibnamefont{Sayama}},
  \bibinfo{author}{\bibfnamefont{K.}~\bibnamefont{Hara}},
  \bibinfo{author}{\bibfnamefont{N.}~\bibnamefont{Mori}},
  \bibinfo{author}{\bibfnamefont{M.}~\bibnamefont{Satsuki}},
  \bibinfo{author}{\bibfnamefont{S.}~\bibnamefont{Suga}},
  \bibinfo{author}{\bibfnamefont{S.}~\bibnamefont{Tsukagoshi}},
  \bibinfo{author}{\bibfnamefont{Y.}~\bibnamefont{Abe}},
  \bibinfo{author}{\bibfnamefont{H.}~\bibnamefont{Sugihara}}, \bibnamefont{and}
  \bibinfo{author}{\bibfnamefont{H.}~\bibnamefont{Arakawa}},
  \bibinfo{journal}{Chem. Commun.} pp. \bibinfo{pages}{1173--1174}
  (\bibinfo{year}{2000}).

\bibitem[{\citenamefont{Yang et~al.}(2012)\citenamefont{Yang, Wang, Duan, Hu,
  Huang, Peng, Huang, and Gong}}]{Yang12}
\bibinfo{author}{\bibfnamefont{T.}~\bibnamefont{Yang}},
  \bibinfo{author}{\bibfnamefont{M.}~\bibnamefont{Wang}},
  \bibinfo{author}{\bibfnamefont{C.}~\bibnamefont{Duan}},
  \bibinfo{author}{\bibfnamefont{X.}~\bibnamefont{Hu}},
  \bibinfo{author}{\bibfnamefont{L.}~\bibnamefont{Huang}},
  \bibinfo{author}{\bibfnamefont{J.}~\bibnamefont{Peng}},
  \bibinfo{author}{\bibfnamefont{F.}~\bibnamefont{Huang}}, \bibnamefont{and}
  \bibinfo{author}{\bibfnamefont{X.}~\bibnamefont{Gong}},
  \bibinfo{journal}{Energy Environ. Sci.} \textbf{\bibinfo{volume}{5}},
  \bibinfo{pages}{8208} (\bibinfo{year}{2012}).

\bibitem[{\citenamefont{Li et~al.}(2012)\citenamefont{Li, Choy, Huo, Xie, Sha,
  Ding, Guo, Li, Hou, You et~al.}}]{Li12}
\bibinfo{author}{\bibfnamefont{X.}~\bibnamefont{Li}},
  \bibinfo{author}{\bibfnamefont{W.~C.~H.} \bibnamefont{Choy}},
  \bibinfo{author}{\bibfnamefont{L.}~\bibnamefont{Huo}},
  \bibinfo{author}{\bibfnamefont{F.}~\bibnamefont{Xie}},
  \bibinfo{author}{\bibfnamefont{W.~E.~I.} \bibnamefont{Sha}},
  \bibinfo{author}{\bibfnamefont{B.}~\bibnamefont{Ding}},
  \bibinfo{author}{\bibfnamefont{X.}~\bibnamefont{Guo}},
  \bibinfo{author}{\bibfnamefont{Y.}~\bibnamefont{Li}},
  \bibinfo{author}{\bibfnamefont{J.}~\bibnamefont{Hou}},
  \bibinfo{author}{\bibfnamefont{J.}~\bibnamefont{You}}, \bibnamefont{et~al.},
  \bibinfo{journal}{Adv. Mat.} \textbf{\bibinfo{volume}{24}},
  \bibinfo{pages}{3046} (\bibinfo{year}{2012}).

\bibitem[{\citenamefont{He et~al.}(2011)\citenamefont{He, Zhong, Huang, Wong,
  Wu, Chen, Su, and Cao}}]{He11}
\bibinfo{author}{\bibfnamefont{Z.}~\bibnamefont{He}},
  \bibinfo{author}{\bibfnamefont{C.}~\bibnamefont{Zhong}},
  \bibinfo{author}{\bibfnamefont{X.}~\bibnamefont{Huang}},
  \bibinfo{author}{\bibfnamefont{W.-Y.} \bibnamefont{Wong}},
  \bibinfo{author}{\bibfnamefont{H.}~\bibnamefont{Wu}},
  \bibinfo{author}{\bibfnamefont{L.}~\bibnamefont{Chen}},
  \bibinfo{author}{\bibfnamefont{S.}~\bibnamefont{Su}}, \bibnamefont{and}
  \bibinfo{author}{\bibfnamefont{Y.}~\bibnamefont{Cao}}, \bibinfo{journal}{Adv.
  Mat.} \textbf{\bibinfo{volume}{23}}, \bibinfo{pages}{4636}
  (\bibinfo{year}{2011}).

\bibitem[{\citenamefont{He et~al.}(2012)\citenamefont{He, Zhong, Su, Xu, Wu,
  and Cao}}]{He12}
\bibinfo{author}{\bibfnamefont{Z.}~\bibnamefont{He}},
  \bibinfo{author}{\bibfnamefont{C.}~\bibnamefont{Zhong}},
  \bibinfo{author}{\bibfnamefont{S.}~\bibnamefont{Su}},
  \bibinfo{author}{\bibfnamefont{M.}~\bibnamefont{Xu}},
  \bibinfo{author}{\bibfnamefont{H.}~\bibnamefont{Wu}}, \bibnamefont{and}
  \bibinfo{author}{\bibfnamefont{Y.}~\bibnamefont{Cao}}, \bibinfo{journal}{Nat.
  Photon} \textbf{\bibinfo{volume}{6}}, \bibinfo{pages}{591}
  (\bibinfo{year}{2012}).

\bibitem[{\citenamefont{Zhang et~al.}(2013)\citenamefont{Zhang, Wang, Xu, Ma,
  Li, and Wang}}]{Zhang13}
\bibinfo{author}{\bibfnamefont{M.}~\bibnamefont{Zhang}},
  \bibinfo{author}{\bibfnamefont{Y.}~\bibnamefont{Wang}},
  \bibinfo{author}{\bibfnamefont{M.}~\bibnamefont{Xu}},
  \bibinfo{author}{\bibfnamefont{W.}~\bibnamefont{Ma}},
  \bibinfo{author}{\bibfnamefont{R.}~\bibnamefont{Li}}, \bibnamefont{and}
  \bibinfo{author}{\bibfnamefont{P.}~\bibnamefont{Wang}},
  \bibinfo{journal}{Energy Environ. Sci.} \textbf{\bibinfo{volume}{6}},
  \bibinfo{pages}{2944} (\bibinfo{year}{2013}).

\bibitem[{\citenamefont{Olivares-Amaya
  et~al.}(2011)\citenamefont{Olivares-Amaya, Amador-Bedolla, Hachmann,
  Atahan-Evrenk, Sanchez-Carrera, Vogt, and Aspuru-Guzik}}]{Olivares11}
\bibinfo{author}{\bibfnamefont{R.}~\bibnamefont{Olivares-Amaya}},
  \bibinfo{author}{\bibfnamefont{C.}~\bibnamefont{Amador-Bedolla}},
  \bibinfo{author}{\bibfnamefont{J.}~\bibnamefont{Hachmann}},
  \bibinfo{author}{\bibfnamefont{S.}~\bibnamefont{Atahan-Evrenk}},
  \bibinfo{author}{\bibfnamefont{R.~S.} \bibnamefont{Sanchez-Carrera}},
  \bibinfo{author}{\bibfnamefont{L.}~\bibnamefont{Vogt}}, \bibnamefont{and}
  \bibinfo{author}{\bibfnamefont{A.}~\bibnamefont{Aspuru-Guzik}},
  \bibinfo{journal}{Energy Environ. Sci.} \textbf{\bibinfo{volume}{4}},
  \bibinfo{pages}{4849} (\bibinfo{year}{2011}).

\bibitem[{\citenamefont{Blouin et~al.}(2008)\citenamefont{Blouin, Michaud,
  Gendron, Wakim, Blair, Neagu-Plesu, Belletete, Durocher, Tao, and
  Leclerc}}]{Blouin08}
\bibinfo{author}{\bibfnamefont{N.}~\bibnamefont{Blouin}},
  \bibinfo{author}{\bibfnamefont{A.}~\bibnamefont{Michaud}},
  \bibinfo{author}{\bibfnamefont{D.}~\bibnamefont{Gendron}},
  \bibinfo{author}{\bibfnamefont{S.}~\bibnamefont{Wakim}},
  \bibinfo{author}{\bibfnamefont{E.}~\bibnamefont{Blair}},
  \bibinfo{author}{\bibfnamefont{R.}~\bibnamefont{Neagu-Plesu}},
  \bibinfo{author}{\bibfnamefont{M.}~\bibnamefont{Belletete}},
  \bibinfo{author}{\bibfnamefont{G.}~\bibnamefont{Durocher}},
  \bibinfo{author}{\bibfnamefont{Y.}~\bibnamefont{Tao}}, \bibnamefont{and}
  \bibinfo{author}{\bibfnamefont{M.}~\bibnamefont{Leclerc}},
  \bibinfo{journal}{J. Am. Chem. Soc.} \textbf{\bibinfo{volume}{130}},
  \bibinfo{pages}{732} (\bibinfo{year}{2008}).

\bibitem[{\citenamefont{Hou et~al.}(2008)\citenamefont{Hou, Chen, Zhang, Li,
  and Yang}}]{Jianhui:08}
\bibinfo{author}{\bibfnamefont{J.}~\bibnamefont{Hou}},
  \bibinfo{author}{\bibfnamefont{H.-Y.} \bibnamefont{Chen}},
  \bibinfo{author}{\bibfnamefont{S.}~\bibnamefont{Zhang}},
  \bibinfo{author}{\bibfnamefont{G.}~\bibnamefont{Li}}, \bibnamefont{and}
  \bibinfo{author}{\bibfnamefont{Y.}~\bibnamefont{Yang}}, \bibinfo{journal}{J.
  Am. Chem. Soc.} \textbf{\bibinfo{volume}{130}}, \bibinfo{pages}{16144}
  (\bibinfo{year}{2008}).

\bibitem[{\citenamefont{Mammo et~al.}(2007)\citenamefont{Mammo, Admassie,
  Gadisa, Zhang, Inganas, and Andersson}}]{Mammo07}
\bibinfo{author}{\bibfnamefont{W.}~\bibnamefont{Mammo}},
  \bibinfo{author}{\bibfnamefont{S.}~\bibnamefont{Admassie}},
  \bibinfo{author}{\bibfnamefont{A.}~\bibnamefont{Gadisa}},
  \bibinfo{author}{\bibfnamefont{F.}~\bibnamefont{Zhang}},
  \bibinfo{author}{\bibfnamefont{O.}~\bibnamefont{Inganas}}, \bibnamefont{and}
  \bibinfo{author}{\bibfnamefont{M.~R.} \bibnamefont{Andersson}},
  \bibinfo{journal}{{Sol. Ener. Mat. Sol. Cel}} \textbf{\bibinfo{volume}{91}},
  \bibinfo{pages}{1010 } (\bibinfo{year}{2007}).

\bibitem[{pro()}]{program}
\emph{\bibinfo{title}{{TURBOMOLE V6.6 2014}, a development of {University of
  Karlsruhe} and {Forschungszentrum Karlsruhe GmbH}, 1989-2007, {TURBOMOLE
  GmbH}, since 2007; available from {\tt http://www.turbomole.com}.}}

\bibitem[{\citenamefont{Lee et~al.}(1988)\citenamefont{Lee, Yang, and
  Parr}}]{Lee88}
\bibinfo{author}{\bibfnamefont{C.}~\bibnamefont{Lee}},
  \bibinfo{author}{\bibfnamefont{W.}~\bibnamefont{Yang}}, \bibnamefont{and}
  \bibinfo{author}{\bibfnamefont{R.~G.} \bibnamefont{Parr}},
  \bibinfo{journal}{Phys. Rev. B} \textbf{\bibinfo{volume}{37}},
  \bibinfo{pages}{785} (\bibinfo{year}{1988}).

\bibitem[{\citenamefont{Becke}(1993)}]{Becke93}
\bibinfo{author}{\bibfnamefont{A.~D.} \bibnamefont{Becke}},
  \bibinfo{journal}{J. Chem. Phys.} \textbf{\bibinfo{volume}{98}},
  \bibinfo{pages}{1372} (\bibinfo{year}{1993}).

\bibitem[{\citenamefont{Weigend et~al.}(1998)\citenamefont{Weigend, Häser,
  Patzelt, and Ahlrichs}}]{Weigend98}
\bibinfo{author}{\bibfnamefont{F.}~\bibnamefont{Weigend}},
  \bibinfo{author}{\bibfnamefont{M.}~\bibnamefont{Häser}},
  \bibinfo{author}{\bibfnamefont{H.}~\bibnamefont{Patzelt}}, \bibnamefont{and}
  \bibinfo{author}{\bibfnamefont{R.}~\bibnamefont{Ahlrichs}},
  \bibinfo{journal}{Chem. Phys. Lett.} \textbf{\bibinfo{volume}{294}},
  \bibinfo{pages}{143 } (\bibinfo{year}{1998}).

\bibitem[{\citenamefont{Casida}(1995)}]{Casida:95}
\bibinfo{author}{\bibfnamefont{M.~E.} \bibnamefont{Casida}},
  \emph{\bibinfo{title}{Recent Advancesin Density Functional Methods}},
  vol.~\bibinfo{volume}{1} (\bibinfo{publisher}{World Scientific: Singapore},
  \bibinfo{year}{1995}).

\bibitem[{\citenamefont{Stephens and Harada}(2010)}]{Stephens10}
\bibinfo{author}{\bibfnamefont{P.~J.} \bibnamefont{Stephens}} \bibnamefont{and}
  \bibinfo{author}{\bibfnamefont{N.}~\bibnamefont{Harada}},
  \bibinfo{journal}{Chirality} \textbf{\bibinfo{volume}{22}},
  \bibinfo{pages}{229} (\bibinfo{year}{2010}).

\bibitem[{\citenamefont{Steffen et~al.}(2010)\citenamefont{Steffen, Thomas,
  Huniar, Hellweg, Rubner, and Schroer}}]{Steffen10}
\bibinfo{author}{\bibfnamefont{C.}~\bibnamefont{Steffen}},
  \bibinfo{author}{\bibfnamefont{K.}~\bibnamefont{Thomas}},
  \bibinfo{author}{\bibfnamefont{U.}~\bibnamefont{Huniar}},
  \bibinfo{author}{\bibfnamefont{A.}~\bibnamefont{Hellweg}},
  \bibinfo{author}{\bibfnamefont{O.}~\bibnamefont{Rubner}}, \bibnamefont{and}
  \bibinfo{author}{\bibfnamefont{A.}~\bibnamefont{Schroer}},
  \bibinfo{journal}{J. Comp. Chem.} \textbf{\bibinfo{volume}{31}},
  \bibinfo{pages}{2967} (\bibinfo{year}{2010}).

\bibitem[{\citenamefont{Kippelen and Bredas}(2009)}]{Kippelen09}
\bibinfo{author}{\bibfnamefont{B.}~\bibnamefont{Kippelen}} \bibnamefont{and}
  \bibinfo{author}{\bibfnamefont{J.-L.} \bibnamefont{Bredas}},
  \bibinfo{journal}{Energy Environ. Sci.} \textbf{\bibinfo{volume}{2}},
  \bibinfo{pages}{251} (\bibinfo{year}{2009}).

\bibitem[{\citenamefont{Qi and Wang}(2013)}]{Qi13}
\bibinfo{author}{\bibfnamefont{B.}~\bibnamefont{Qi}} \bibnamefont{and}
  \bibinfo{author}{\bibfnamefont{J.}~\bibnamefont{Wang}},
  \bibinfo{journal}{Phys. Chem. Chem. Phys.} \textbf{\bibinfo{volume}{15}},
  \bibinfo{pages}{8972} (\bibinfo{year}{2013}).

\bibitem[{\citenamefont{Scharber et~al.}(2006)\citenamefont{Scharber,
  Mühlbacher, Koppe, Denk, Waldauf, Heeger, and Brabec}}]{Scharber06}
\bibinfo{author}{\bibfnamefont{M.~C.} \bibnamefont{Scharber}},
  \bibinfo{author}{\bibfnamefont{D.}~\bibnamefont{Mühlbacher}},
  \bibinfo{author}{\bibfnamefont{M.}~\bibnamefont{Koppe}},
  \bibinfo{author}{\bibfnamefont{P.}~\bibnamefont{Denk}},
  \bibinfo{author}{\bibfnamefont{C.}~\bibnamefont{Waldauf}},
  \bibinfo{author}{\bibfnamefont{A.~J.} \bibnamefont{Heeger}},
  \bibnamefont{and} \bibinfo{author}{\bibfnamefont{C.~J.}
  \bibnamefont{Brabec}}, \bibinfo{journal}{Adv. Mat.}
  \textbf{\bibinfo{volume}{18}}, \bibinfo{pages}{789} (\bibinfo{year}{2006}).

\bibitem[{\citenamefont{Mühlbacher et~al.}(2006)\citenamefont{Mühlbacher,
  Scharber, Morana, Zhu, Waller, Gaudiana, and Brabec}}]{Muhlbacher06}
\bibinfo{author}{\bibfnamefont{D.}~\bibnamefont{Mühlbacher}},
  \bibinfo{author}{\bibfnamefont{M.}~\bibnamefont{Scharber}},
  \bibinfo{author}{\bibfnamefont{M.}~\bibnamefont{Morana}},
  \bibinfo{author}{\bibfnamefont{Z.}~\bibnamefont{Zhu}},
  \bibinfo{author}{\bibfnamefont{D.}~\bibnamefont{Waller}},
  \bibinfo{author}{\bibfnamefont{R.}~\bibnamefont{Gaudiana}}, \bibnamefont{and}
  \bibinfo{author}{\bibfnamefont{C.}~\bibnamefont{Brabec}},
  \bibinfo{journal}{Adv. Mat.} \textbf{\bibinfo{volume}{18}},
  \bibinfo{pages}{2884} (\bibinfo{year}{2006}).

\bibitem[{\citenamefont{Wang et~al.}(2014)\citenamefont{Wang, Zhang, Ding,
  Zhao, and Geng}}]{Dongmei14}
\bibinfo{author}{\bibfnamefont{D.}~\bibnamefont{Wang}},
  \bibinfo{author}{\bibfnamefont{X.}~\bibnamefont{Zhang}},
  \bibinfo{author}{\bibfnamefont{W.}~\bibnamefont{Ding}},
  \bibinfo{author}{\bibfnamefont{X.}~\bibnamefont{Zhao}}, \bibnamefont{and}
  \bibinfo{author}{\bibfnamefont{Z.}~\bibnamefont{Geng}},
  \bibinfo{journal}{Comp. Theo. Chem.} \textbf{\bibinfo{volume}{1029}},
  \bibinfo{pages}{68 } (\bibinfo{year}{2014}).

\bibitem[{\citenamefont{Bourass et~al.}(2016)\citenamefont{Bourass, Benjelloun,
  Hamidi, Benzakour, Mcharfi, Sfaira, Serein-Spirau, Lère-Porte, Sotiropoulos,
  Bouzzine et~al.}}]{Bourass13}
\bibinfo{author}{\bibfnamefont{M.}~\bibnamefont{Bourass}},
  \bibinfo{author}{\bibfnamefont{A.~T.} \bibnamefont{Benjelloun}},
  \bibinfo{author}{\bibfnamefont{M.}~\bibnamefont{Hamidi}},
  \bibinfo{author}{\bibfnamefont{M.}~\bibnamefont{Benzakour}},
  \bibinfo{author}{\bibfnamefont{M.}~\bibnamefont{Mcharfi}},
  \bibinfo{author}{\bibfnamefont{M.}~\bibnamefont{Sfaira}},
  \bibinfo{author}{\bibfnamefont{F.}~\bibnamefont{Serein-Spirau}},
  \bibinfo{author}{\bibfnamefont{J.-P.} \bibnamefont{Lère-Porte}},
  \bibinfo{author}{\bibfnamefont{J.-M.} \bibnamefont{Sotiropoulos}},
  \bibinfo{author}{\bibfnamefont{S.~M.} \bibnamefont{Bouzzine}},
  \bibnamefont{et~al.}, \bibinfo{journal}{J. Saudi Chem. Soc.}
  \textbf{\bibinfo{volume}{20}}, \bibinfo{pages}{S415} (\bibinfo{year}{2016}).

\bibitem[{\citenamefont{Ding et~al.}(2013)\citenamefont{Ding, Wang, Geng, Zhao,
  and Xu}}]{Ding13}
\bibinfo{author}{\bibfnamefont{W.-L.} \bibnamefont{Ding}},
  \bibinfo{author}{\bibfnamefont{D.-M.} \bibnamefont{Wang}},
  \bibinfo{author}{\bibfnamefont{Z.-Y.} \bibnamefont{Geng}},
  \bibinfo{author}{\bibfnamefont{X.-L.} \bibnamefont{Zhao}}, \bibnamefont{and}
  \bibinfo{author}{\bibfnamefont{W.-B.} \bibnamefont{Xu}},
  \bibinfo{journal}{Dyes Pigm.} \textbf{\bibinfo{volume}{98}},
  \bibinfo{pages}{125 } (\bibinfo{year}{2013}).

\bibitem[{\citenamefont{Zhang et~al.}(2012{\natexlab{a}})\citenamefont{Zhang,
  Kan, Li, Geng, Wu, and Su}}]{Zhang12}
\bibinfo{author}{\bibfnamefont{J.}~\bibnamefont{Zhang}},
  \bibinfo{author}{\bibfnamefont{Y.-H.} \bibnamefont{Kan}},
  \bibinfo{author}{\bibfnamefont{H.-B.} \bibnamefont{Li}},
  \bibinfo{author}{\bibfnamefont{Y.}~\bibnamefont{Geng}},
  \bibinfo{author}{\bibfnamefont{Y.}~\bibnamefont{Wu}}, \bibnamefont{and}
  \bibinfo{author}{\bibfnamefont{Z.-M.} \bibnamefont{Su}},
  \bibinfo{journal}{Dyes Pigm.} \textbf{\bibinfo{volume}{95}},
  \bibinfo{pages}{313 } (\bibinfo{year}{2012}{\natexlab{a}}).

\bibitem[{\citenamefont{Fan et~al.}(2012)\citenamefont{Fan, Tan, and
  Deng}}]{Fan12}
\bibinfo{author}{\bibfnamefont{W.}~\bibnamefont{Fan}},
  \bibinfo{author}{\bibfnamefont{D.}~\bibnamefont{Tan}}, \bibnamefont{and}
  \bibinfo{author}{\bibfnamefont{W.-Q.} \bibnamefont{Deng}},
  \bibinfo{journal}{Chem. Phys. Chem} \textbf{\bibinfo{volume}{13}},
  \bibinfo{pages}{1966} (\bibinfo{year}{2012}).

\bibitem[{\citenamefont{Zhang et~al.}(2012{\natexlab{b}})\citenamefont{Zhang,
  Li, Sun, Geng, Wu, and Su}}]{Zhang12-1}
\bibinfo{author}{\bibfnamefont{J.}~\bibnamefont{Zhang}},
  \bibinfo{author}{\bibfnamefont{H.-B.} \bibnamefont{Li}},
  \bibinfo{author}{\bibfnamefont{S.-L.} \bibnamefont{Sun}},
  \bibinfo{author}{\bibfnamefont{Y.}~\bibnamefont{Geng}},
  \bibinfo{author}{\bibfnamefont{Y.}~\bibnamefont{Wu}}, \bibnamefont{and}
  \bibinfo{author}{\bibfnamefont{Z.-M.} \bibnamefont{Su}}, \bibinfo{journal}{J.
  Mater. Chem.} \textbf{\bibinfo{volume}{22}}, \bibinfo{pages}{568}
  (\bibinfo{year}{2012}{\natexlab{b}}).

\bibitem[{\citenamefont{Wang et~al.}(2011)\citenamefont{Wang, Bai, Xia, Feng,
  Zhang, and Pan}}]{Wang11}
\bibinfo{author}{\bibfnamefont{J.}~\bibnamefont{Wang}},
  \bibinfo{author}{\bibfnamefont{F.-Q.} \bibnamefont{Bai}},
  \bibinfo{author}{\bibfnamefont{B.-H.} \bibnamefont{Xia}},
  \bibinfo{author}{\bibfnamefont{L.}~\bibnamefont{Feng}},
  \bibinfo{author}{\bibfnamefont{H.-X.} \bibnamefont{Zhang}}, \bibnamefont{and}
  \bibinfo{author}{\bibfnamefont{Q.-J.} \bibnamefont{Pan}},
  \bibinfo{journal}{Phys. Chem. Chem. Phys.} \textbf{\bibinfo{volume}{13}},
  \bibinfo{pages}{2206} (\bibinfo{year}{2011}).

\bibitem[{\citenamefont{Sang-aroon et~al.}(2012)\citenamefont{Sang-aroon,
  Saekow, and Amornkitbamrung}}]{Wichien12}
\bibinfo{author}{\bibfnamefont{W.}~\bibnamefont{Sang-aroon}},
  \bibinfo{author}{\bibfnamefont{S.}~\bibnamefont{Saekow}}, \bibnamefont{and}
  \bibinfo{author}{\bibfnamefont{V.}~\bibnamefont{Amornkitbamrung}},
  \bibinfo{journal}{J. Photochem. Photobiol.} \textbf{\bibinfo{volume}{236}},
  \bibinfo{pages}{35 } (\bibinfo{year}{2012}).

\bibitem[{\citenamefont{Katoh et~al.}(2004)\citenamefont{Katoh, Furube,
  Yoshihara, Hara, Fujihashi, Takano, Murata, Arakawa, , and
  Tachiya}}]{Ryuzi04}
\bibinfo{author}{\bibfnamefont{R.}~\bibnamefont{Katoh}},
  \bibinfo{author}{\bibfnamefont{A.}~\bibnamefont{Furube}},
  \bibinfo{author}{\bibfnamefont{T.}~\bibnamefont{Yoshihara}},
  \bibinfo{author}{\bibfnamefont{K.}~\bibnamefont{Hara}},
  \bibinfo{author}{\bibfnamefont{G.}~\bibnamefont{Fujihashi}},
  \bibinfo{author}{\bibfnamefont{S.}~\bibnamefont{Takano}},
  \bibinfo{author}{\bibfnamefont{S.}~\bibnamefont{Murata}},
  \bibinfo{author}{\bibfnamefont{H.}~\bibnamefont{Arakawa}}, ,
  \bibnamefont{and} \bibinfo{author}{\bibfnamefont{M.}~\bibnamefont{Tachiya}},
  \bibinfo{journal}{J. Phys. Chem. B} \textbf{\bibinfo{volume}{108}},
  \bibinfo{pages}{4818} (\bibinfo{year}{2004}).

\bibitem[{\citenamefont{{J. P. Perdew and K. Burke and M.
  Ernzerhof}}(1996)}]{Perdew:96}
\bibinfo{author}{\bibnamefont{{J. P. Perdew and K. Burke and M. Ernzerhof}}},
  \bibinfo{journal}{{Phys. Rev. Lett.}} \textbf{\bibinfo{volume}{77}},
  \bibinfo{pages}{3865} (\bibinfo{year}{1996}).

\bibitem[{\citenamefont{Lewars}(2011)}]{Lewars11}
\bibinfo{author}{\bibfnamefont{E.}~\bibnamefont{Lewars}},
  \emph{\bibinfo{title}{Computational Chemistry: Introduction to the Theory and
  Applications of Molecular and Quantum Mechanics}}
  (\bibinfo{publisher}{Springer Netherlands}, \bibinfo{year}{2011}),
  \bibinfo{edition}{2nd} ed.

\bibitem[{\citenamefont{Zhang et~al.}(2010)\citenamefont{Zhang, Du, Sun, and
  Sun}}]{Zhang10}
\bibinfo{author}{\bibfnamefont{R.}~\bibnamefont{Zhang}},
  \bibinfo{author}{\bibfnamefont{B.}~\bibnamefont{Du}},
  \bibinfo{author}{\bibfnamefont{G.}~\bibnamefont{Sun}}, \bibnamefont{and}
  \bibinfo{author}{\bibfnamefont{Y.}~\bibnamefont{Sun}},
  \bibinfo{journal}{Spectrochim. Acta Mol. Biomol. Spectrosc.}
  \textbf{\bibinfo{volume}{75}}, \bibinfo{pages}{1115 } (\bibinfo{year}{2010}).

\bibitem[{\citenamefont{Kosar and Albayrak}(2011)}]{Kosar11}
\bibinfo{author}{\bibfnamefont{B.}~\bibnamefont{Kosar}} \bibnamefont{and}
  \bibinfo{author}{\bibfnamefont{C.}~\bibnamefont{Albayrak}},
  \bibinfo{journal}{Spectrochim. Acta Mol. Biomol. Spectrosc.}
  \textbf{\bibinfo{volume}{78}}, \bibinfo{pages}{160 } (\bibinfo{year}{2011}).

\bibitem[{\citenamefont{Mohr et~al.}(2015)\citenamefont{Mohr, Aroulmoji,
  Ravindran, Müller, Ranjitha, Rajarajan, and Anbarasan}}]{Mohr15}
\bibinfo{author}{\bibfnamefont{T.}~\bibnamefont{Mohr}},
  \bibinfo{author}{\bibfnamefont{V.}~\bibnamefont{Aroulmoji}},
  \bibinfo{author}{\bibfnamefont{R.~S.} \bibnamefont{Ravindran}},
  \bibinfo{author}{\bibfnamefont{M.}~\bibnamefont{Müller}},
  \bibinfo{author}{\bibfnamefont{S.}~\bibnamefont{Ranjitha}},
  \bibinfo{author}{\bibfnamefont{G.}~\bibnamefont{Rajarajan}},
  \bibnamefont{and}
  \bibinfo{author}{\bibfnamefont{P.}~\bibnamefont{Anbarasan}},
  \bibinfo{journal}{Spectrochim. Acta Mol. Biomol. Spectrosc.}
  \textbf{\bibinfo{volume}{135}}, \bibinfo{pages}{1066 }
  (\bibinfo{year}{2015}).

\bibitem[{\citenamefont{Xie et~al.}(2015)\citenamefont{Xie, Wang, Xia, Bai,
  Jia, Rim, and Zhang}}]{Xie15}
\bibinfo{author}{\bibfnamefont{M.}~\bibnamefont{Xie}},
  \bibinfo{author}{\bibfnamefont{J.}~\bibnamefont{Wang}},
  \bibinfo{author}{\bibfnamefont{H.-Q.} \bibnamefont{Xia}},
  \bibinfo{author}{\bibfnamefont{F.-Q.} \bibnamefont{Bai}},
  \bibinfo{author}{\bibfnamefont{R.}~\bibnamefont{Jia}},
  \bibinfo{author}{\bibfnamefont{J.-G.} \bibnamefont{Rim}}, \bibnamefont{and}
  \bibinfo{author}{\bibfnamefont{H.-X.} \bibnamefont{Zhang}},
  \bibinfo{journal}{RSC Adv.} \textbf{\bibinfo{volume}{5}},
  \bibinfo{pages}{33653} (\bibinfo{year}{2015}).

\bibitem[{\citenamefont{Chen and Cao}(2009)}]{Chen09}
\bibinfo{author}{\bibfnamefont{J.}~\bibnamefont{Chen}} \bibnamefont{and}
  \bibinfo{author}{\bibfnamefont{Y.}~\bibnamefont{Cao}}, \bibinfo{journal}{Acc.
  Chem. Res.} \textbf{\bibinfo{volume}{42}}, \bibinfo{pages}{1709}
  (\bibinfo{year}{2009}).

\bibitem[{\citenamefont{Mondal et~al.}(2009)\citenamefont{Mondal, Miyaki,
  Becerril, Norton, Parmer, Mayer, Tang, Brédas, McGehee, and Bao}}]{Mondal09}
\bibinfo{author}{\bibfnamefont{R.}~\bibnamefont{Mondal}},
  \bibinfo{author}{\bibfnamefont{N.}~\bibnamefont{Miyaki}},
  \bibinfo{author}{\bibfnamefont{H.~A.} \bibnamefont{Becerril}},
  \bibinfo{author}{\bibfnamefont{J.~E.} \bibnamefont{Norton}},
  \bibinfo{author}{\bibfnamefont{J.}~\bibnamefont{Parmer}},
  \bibinfo{author}{\bibfnamefont{A.~C.} \bibnamefont{Mayer}},
  \bibinfo{author}{\bibfnamefont{M.~L.} \bibnamefont{Tang}},
  \bibinfo{author}{\bibfnamefont{J.-L.} \bibnamefont{Brédas}},
  \bibinfo{author}{\bibfnamefont{M.~D.} \bibnamefont{McGehee}},
  \bibnamefont{and} \bibinfo{author}{\bibfnamefont{Z.}~\bibnamefont{Bao}},
  \bibinfo{journal}{Chem. Mat.} \textbf{\bibinfo{volume}{21}},
  \bibinfo{pages}{3618} (\bibinfo{year}{2009}).

\bibitem[{\citenamefont{Wang et~al.}(2008)\citenamefont{Wang, Wang, Lan, Luo,
  Zhuang, Peng, and Cao}}]{Wang:08}
\bibinfo{author}{\bibfnamefont{E.}~\bibnamefont{Wang}},
  \bibinfo{author}{\bibfnamefont{L.}~\bibnamefont{Wang}},
  \bibinfo{author}{\bibfnamefont{L.}~\bibnamefont{Lan}},
  \bibinfo{author}{\bibfnamefont{C.}~\bibnamefont{Luo}},
  \bibinfo{author}{\bibfnamefont{W.}~\bibnamefont{Zhuang}},
  \bibinfo{author}{\bibfnamefont{J.}~\bibnamefont{Peng}}, \bibnamefont{and}
  \bibinfo{author}{\bibfnamefont{Y.}~\bibnamefont{Cao}}, \bibinfo{journal}{App.
  Phys. Lett.} \textbf{\bibinfo{volume}{92}}, \bibinfo{pages}{033307}
  (\bibinfo{year}{2008}).

\bibitem[{\citenamefont{Yuen et~al.}(2010)\citenamefont{Yuen, Hor, Jovanovic,
  Preston, Klenkler, Bamsey, and Loutfy}}]{Yuen10}
\bibinfo{author}{\bibfnamefont{A.~P.} \bibnamefont{Yuen}},
  \bibinfo{author}{\bibfnamefont{A.-M.} \bibnamefont{Hor}},
  \bibinfo{author}{\bibfnamefont{S.~M.} \bibnamefont{Jovanovic}},
  \bibinfo{author}{\bibfnamefont{J.~S.} \bibnamefont{Preston}},
  \bibinfo{author}{\bibfnamefont{R.~A.} \bibnamefont{Klenkler}},
  \bibinfo{author}{\bibfnamefont{N.~M.} \bibnamefont{Bamsey}},
  \bibnamefont{and} \bibinfo{author}{\bibfnamefont{R.~O.}
  \bibnamefont{Loutfy}}, \bibinfo{journal}{Sol. Energ. Mat. Sol. Cells}
  \textbf{\bibinfo{volume}{94}}, \bibinfo{pages}{2455 } (\bibinfo{year}{2010}).

\bibitem[{\citenamefont{Horie et~al.}(2012)\citenamefont{Horie, Kettle, Yu,
  Majewski, Chang, Kirkpatrick, Tuladhar, Nelson, Saunders, and
  Turner}}]{Horie:12}
\bibinfo{author}{\bibfnamefont{M.}~\bibnamefont{Horie}},
  \bibinfo{author}{\bibfnamefont{J.}~\bibnamefont{Kettle}},
  \bibinfo{author}{\bibfnamefont{C.-Y.} \bibnamefont{Yu}},
  \bibinfo{author}{\bibfnamefont{L.~A.} \bibnamefont{Majewski}},
  \bibinfo{author}{\bibfnamefont{S.-W.} \bibnamefont{Chang}},
  \bibinfo{author}{\bibfnamefont{J.}~\bibnamefont{Kirkpatrick}},
  \bibinfo{author}{\bibfnamefont{S.~M.} \bibnamefont{Tuladhar}},
  \bibinfo{author}{\bibfnamefont{J.}~\bibnamefont{Nelson}},
  \bibinfo{author}{\bibfnamefont{B.~R.} \bibnamefont{Saunders}},
  \bibnamefont{and} \bibinfo{author}{\bibfnamefont{M.~L.}
  \bibnamefont{Turner}}, \bibinfo{journal}{J. Mater. Chem.}
  \textbf{\bibinfo{volume}{22}}, \bibinfo{pages}{381} (\bibinfo{year}{2012}).

\bibitem[{\citenamefont{Muñiz et~al.}(2013)\citenamefont{Muñiz, Sansores, and
  Castillo}}]{Muniz13}
\bibinfo{author}{\bibfnamefont{J.}~\bibnamefont{Muñiz}},
  \bibinfo{author}{\bibfnamefont{E.}~\bibnamefont{Sansores}}, \bibnamefont{and}
  \bibinfo{author}{\bibfnamefont{R.}~\bibnamefont{Castillo}},
  \bibinfo{journal}{Theor. Chem. Acc.} \textbf{\bibinfo{volume}{132}},
  \bibinfo{pages}{1373} (\bibinfo{year}{2013}).

\bibitem[{\citenamefont{Asiri et~al.}(2011)\citenamefont{Asiri, Karabacak,
  Kurt, and Alamry}}]{Asiri20}
\bibinfo{author}{\bibfnamefont{A.~M.} \bibnamefont{Asiri}},
  \bibinfo{author}{\bibfnamefont{M.}~\bibnamefont{Karabacak}},
  \bibinfo{author}{\bibfnamefont{M.}~\bibnamefont{Kurt}}, \bibnamefont{and}
  \bibinfo{author}{\bibfnamefont{K.~A.} \bibnamefont{Alamry}},
  \bibinfo{journal}{Spectrochim. Acta Mol. Biomol. Spectrosc.}
  \textbf{\bibinfo{volume}{82}}, \bibinfo{pages}{444 } (\bibinfo{year}{2011}).

\bibitem[{\citenamefont{Senge et~al.}(2007)\citenamefont{Senge, Fazekas,
  Notaras, Blau, Zawadzka, Locos, and Ni~Mhuircheartaigh}}]{Senge07}
\bibinfo{author}{\bibfnamefont{M.}~\bibnamefont{Senge}},
  \bibinfo{author}{\bibfnamefont{M.}~\bibnamefont{Fazekas}},
  \bibinfo{author}{\bibfnamefont{E.}~\bibnamefont{Notaras}},
  \bibinfo{author}{\bibfnamefont{W.}~\bibnamefont{Blau}},
  \bibinfo{author}{\bibfnamefont{M.}~\bibnamefont{Zawadzka}},
  \bibinfo{author}{\bibfnamefont{O.}~\bibnamefont{Locos}}, \bibnamefont{and}
  \bibinfo{author}{\bibfnamefont{E.}~\bibnamefont{Ni~Mhuircheartaigh}},
  \bibinfo{journal}{Adv. Mat.} \textbf{\bibinfo{volume}{19}},
  \bibinfo{pages}{2737} (\bibinfo{year}{2007}).

\bibitem[{\citenamefont{{Martin Gouterman}}(1961)}]{Gouterman:61}
\bibinfo{author}{\bibnamefont{{Martin Gouterman}}}, \bibinfo{journal}{{J. Mol.
  Spectrosc.}} \textbf{\bibinfo{volume}{6}}, \bibinfo{pages}{138}
  (\bibinfo{year}{1961}).

\bibitem[{\citenamefont{{C.R. Wronski and J.M. Pearce and J. Deng and V. Vlahos
  and R.W. Collins}}(2004)}]{Wronski:04}
\bibinfo{author}{\bibnamefont{{C.R. Wronski and J.M. Pearce and J. Deng and V.
  Vlahos and R.W. Collins}}}, \bibinfo{journal}{{Thin Solid Films}}
  \textbf{\bibinfo{volume}{451-452}}, \bibinfo{pages}{470}
  (\bibinfo{year}{2004}).

\bibitem[{\citenamefont{{Licht, S. and Wang, B. and Soga, T. and Umeno,
  M.}}(1999)}]{Soga:99}
\bibinfo{author}{\bibnamefont{{Licht, S. and Wang, B. and Soga, T. and Umeno,
  M.}}}, \bibinfo{journal}{{App. Phys. Lett.}} \textbf{\bibinfo{volume}{74}},
  \bibinfo{pages}{4055} (\bibinfo{year}{1999}).

\end{thebibliography}

%
%


\clearpage
\section*{Figures}


\clearpage
\section*{TABLES}

\begin{longtable}{|c|c|c|c|c|c|}
\hline
\multirow{4}{1.7cm}{\centering Molecular system} & \multirow{4}{1.8cm}{\centering One-electron transition} & \multirow{4}{2cm}{\centering Absortion wavelength   $\lambda_1$, $\lambda_2$  ($\lambda_{Abs}$ nm)} & \multirow{4}{1.9cm}{\centering Oscillator strength ($f$)} & \multirow{4}{2cm}{\centering Excitation energy E$_{ex}$ (eV) } & \multirow{4}{2.6cm}{\centering $|$CI$|$-coefficient($\%$)}\\
&&&&&\\
&&&&&\\
&&&&&\\
\hline
\endfirsthead
\hline
\multirow{4}{1.7cm}{\centering Molecular system} & \multirow{4}{1.8cm}{\centering One-electron transition} & \multirow{4}{2cm}{\centering Absortion wavelength   $\lambda_1$, $\lambda_2$  ($\lambda_{Abs}$ nm)} & \multirow{4}{1.9cm}{\centering Oscillator strength ($f$)} & \multirow{4}{2cm}{\centering Excitation energy E$_{ex}$ (eV) } & \multirow{4}{2.6cm}{\centering $|$CI$|$-coefficient ($\%$)}\\
&&&&&\\
&&&&&\\
&&&&&\\
\hline
\endhead

\endfoot
\endlastfoot

\multirow{2}{1.7cm}{\centering	 S16	}	&		$S_0 \rightarrow S_1$		&		611.6		&		0.573		&		2.028		&		$ H \rightarrow L $	(	97.8	)		\\	
			&		$ S_0 \rightarrow S_3$		&		470.7		&		0.960		&		2.634		&		H $\rightarrow $ L+1	(	95.8	)		\\	\hline
\multirow{2}{1.7cm}{\centering	 S17	}	&		$S_0 \rightarrow S_1$		&		641.9		&		0.584		&		1.932		&		$ H \rightarrow L $	(	97.5	)		\\	
			&		$ S_0 \rightarrow S_3$		&		407.0		&		1.058		&		3.046		&		H $ \rightarrow $ L+1	(	94.2	)		\\	\hline
\multirow{2}{1.7cm}{\centering	 S19	}	&		$S_0 \rightarrow S_1$		&		550.4		&		0.555		&		2.253		&		$ H \rightarrow L $	(	96.9	)		\\	
			&		$ S_0 \rightarrow S_4$		&		389.2		&		0.755		&		3.186		&		H $ \rightarrow $ L+1	(	94.9	)		\\	\hline
\multirow{2}{1.7cm}{\centering	 S20	}	&		$S_0 \rightarrow S_1$		&		583.2		&		0.573		&		2.126		&		$ H \rightarrow L $	(	98.1	)		\\	
			&		$ S_0 \rightarrow S_4$	&		377.1		&		1.059		&		3.288		&		H $ \rightarrow $ L+1	(	96	)		\\	\hline
\multirow{2}{1.7cm}{\centering S21	}	&		$S_0 \rightarrow S_1$		&		428.9		&		0.375		&		1.335		&		$ H \rightarrow L $	(	98.7	)		\\	
			&		$ S_0 \rightarrow S_2$		&		595.9		&		0.584		&		2.081		&		H $ \rightarrow $ L+1	(	89.1	)		\\	\hline
\multirow{2}{1.7cm}{\centering	 S33	}	& $S_0 \rightarrow S_1$	& 612.2	 & 1.282	&	2.025	& $ H \rightarrow L $	(98.9)	\\	
			&	$S_0 \rightarrow S_3$	&	408.3	&	0.315 &	3.036 &	H $ \rightarrow $ L+2 (12)	\\	\hline
\multirow{2}{1.7cm}{\centering	 S46	}	& $S_0 \rightarrow S_1$	 & 580.7 & 1.647	&	2.135 &  $ H \rightarrow L $	(98.4) \\	
 
 & $S_0 \rightarrow S_6$ & 356.5 & 0.517	& 3.477 &	 H-1 $\rightarrow $ L+1 (73.5)\\	\hline

\multirow{2}{1.7cm}{\centering	S47	}	& $S_0 \rightarrow S_1$	 & 634.9 & 2.563	&	1.953	&  $ H \rightarrow L $	(97.6)	\\	
	
& $S_0 \rightarrow S_6 $ & 398.3 &	0.65	&	3.113 & H-1 $\rightarrow $ L+1 (49.5 )\\	\hline

\multirow{2}{1.7cm}{\centering	 S49	}	& $S_0 \rightarrow S_1$		& 576.5	 &	1.629	& 2.151	& $ H \rightarrow L $ (98.5)	\\	

& $S_0 \rightarrow S_7$ &353.1	&	0.437 & 3.511	&  H-1 $\rightarrow $ L+1 (78) \\	\hline

\multirow{2}{1.7cm}{\centering	S50	}	& $S_0 \rightarrow S_1$	& 576.5	 & 1.665	&	2.151 &  $ H \rightarrow L $ (98.4)	\\	

& $S_0 \rightarrow S_7$ & 355.5	& 0.5 &	3.48	& H-1 $\rightarrow $ L+1 (61)	\\	\hline

\caption{\label{estado_exitado}Excited state properties of the selected systems. Note that HOMO and LUMO correspond to H and L, respectively.}
\end{longtable}

\end{document}